\documentclass[12pt]{article}
\usepackage[utf8]{inputenc}
\usepackage{graphicx}
\usepackage{tikz}
\usepackage{hyperref}
\usepackage{xcolor}
\usepackage{array}
\usepackage{bm}
\hypersetup{
    colorlinks = true,
    urlcolor   = blue,
    citecolor  = black,
}

\newcommand{\RomanNumeralCaps}[1]
\linenumbers

\usepackage{verbatim}
\usepackage{amsmath}
\usepackage{amssymb}
\numberwithin{equation}{section}

\title{Explicit formulas for forced convection in a shrouded longitudinal-fin heat sink with clearance}
\author{Toby L. Kirk\footnotemark[2], Marc Hodes\footnotemark[4]}
\date{\footnotemark[2]~\it{School of Mathematical Sciences, University of Southampton\\
\vspace{0.15cm}
Southampton, SO14 3ZH, UK}\\ 
\vspace{0.3cm}
\footnotemark[4]~\it{Department of Mechanical Engineering, Tufts University\\ 
\vspace{0.15cm}
Medford MA, 02155, US
}}
\begin{document}

\maketitle

\begin{abstract}
    We consider laminar forced convection in a shrouded longitudinal-fin heat sink (LFHS) with tip clearance, as described by the pioneering study of Sparrow, Baliga and Patankar [1978, \emph{J. Heat Trans}, \textbf{100}(4)]. The base of the LFHS is isothermal but the fins, while thin, are not isothermal, i.e., the conjugate heat transfer problem is of interest. Whereas Sparrow et al. solved the fully developed flow and thermal problems numerically for a range of geometries and fin conductivities, we consider here the physically realistic asymptotic limit where the fins are closely spaced, i.e. the spacing is small relative to their height and the clearance above them. The flow problem in this limit was considered by Miyoshi \emph{et al.} [2024, \emph{J. Fluid Mech}, \textbf{991}, A2], and here we consider the corresponding thermal problem. Using matched asymptotic expansions, we find explicit solutions for the temperature field (in both the fluid and fins) and conjugate Nusselt numbers (local and average). The structure of the asymptotic solutions provides insight into the results of Sparrow \emph{et al.}: the flow is highest in the gap above the fins, hence heat transfer predominantly occurs close to the fin tips.
    The new formulas are compared to numerical solutions and are found to be accurate for practical LFHSs. Significantly, existing analytical results for ducts are for boundaries that are either wholly isothermal, wholly isoflux or with one of these conditions on each wall. Consequently, this study provides the first analytical results for conjugate Nusselt numbers for flow through ducts.
\end{abstract}

\section{Introduction}
Heat sinks are ubiquitous in modern computing and telecommunications hardware. More generally, they are an enabling technology in the thermal management of all electronics and elsewhere. When the flow is unidirectional, generally the fins on them are nearly rectangular in cross-section, in which case the heat sink is referred to
as a longitudinal-fin heat sink (LFHS). LFHSs are manufactured by 
various methods (extrusion, skiving, machining, etc.), each imposing constraints on, e.g., minimum fin spacing and thickness, maximum fin height-to-spacing ratio, materials and cost~\cite{Iyengar-07}. Both air-cooled LFHSs (say, in a laptop) and 
water-cooled ones (say, in a ``cold plate'' attached to a central processing unit, CPU, in a server blade) are common. The materials of LFHSs are most commonly aluminum or copper, although the former is incompatible with water. 

\begin{figure}
    \centering
    \includegraphics[width=0.7\textwidth]{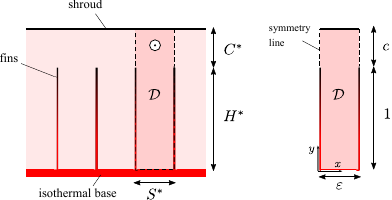}
    \caption{Schematic cross-section of the periodic fin array (left) considered here and by ~\cite{Sparrow1978}; and the dimensionless single period domain $\mathcal{D}$ (right).}
    \label{fig:domain_schematic}
\end{figure}

The foundational study on laminar (forced) convection in an LFHS was published by ~\cite{Sparrow1978}. They considered fully-developed flow and heat transfer and allowed for tip clearance between the top of the fins and an adiabatic shroud, but not for bypass flow around the sides of the LFHS. Viscous dissipation was assumed negligible and thermophysical properties were considered to be constant. Their key results pertained to an isothermal base, a valid assumption in modern applications when, as is common, the fins are attached to a vapor chamber. A key assumption invoked by~\cite{Sparrow1978} was that, geometrically, the fins were considered to be vanishingly thin, and they neglected heat sink edge effects. Consequently, the fluid domain in one period was rectangular and, due to symmetry, its width was half the fin spacing as per Fig.~\ref{fig:domain_schematic}. The dimensional fin spacing was $S^*$, the fin height $H^*$, and the clearance $C^*$. Their hydrodynamic results were provided via tabulations of the Poiseuille number (Po, or product of the friction factor and Reynolds number) as a function of the ratio of the fin-spacing-to-height ratio ($\varepsilon = S^*/H^*$) and fin-clearance-to-height ratio ($c = C^*/H^*$).

In the thermal problem,~\cite{Sparrow1978} assumed that the Biot number based on the thickness of the fins was small; therefore, temperature varied only along its height. Importantly, the conjugate thermal problem was solved by matching temperature and balancing the net heat conduction rate along a differential height of the fin with the heat rate into the fluid along the fin--fluid interface. In addition, heat transfer between the prime surface (between fins) and the fluid was captured. The local Nusselt numbers ($\mathrm{Nu}$) along the fin and prime surface were provided as a function of nondimensional distance $x$ (along the prime surface) or $y$ (along the fin), $\varepsilon$, $c$ and a dimensionless fin conduction parameter $\Omega$ defined as 
\begin{equation}
\Omega = \frac{k_{\mathrm{f}}}{k}\frac{t^*}{2H^*}, \label{eq:Omega}
\end{equation}
where $k_{\mathrm{f}}$ is the thermal conductivity of the fin and $k$ is that of the fluid, with $t^*$ the thickness of the fin. The fin becomes isothermal as $\Omega \rightarrow \infty$. They further provide a Nusselt number averaged over the prime surface and fin, $\overline{\mathrm{Nu}}$, that is independent of $x$ and $y$. 
A key conclusion of~\cite{Sparrow1978} was that the ubiquitous assumption of a constant heat transfer coefficient along the prime surface and fin is generally invalid. Alas, it remains common today. In a subsequent and related study,~\cite{Sparrow1981} relaxed the assumptions of the fins being vanishingly thin in the hydrodynamic problem and the fin temperature being constant across its width. (Axial conduction remains neglected in the fin and the fluid as in our own analysis to follow.) The effects on the Poiseuille number were modest in the parametric ranges considered for realistic heat sink geometries. Moreover, it was shown that the adiabatic fin tip assumption compares well to the case where convection from the fin tip is captured because this assumption causes a marked increase in heat transfer near the tip.

\cite{Karamanis-16}, albeit restricting their attention to fully-shrouded LFHSs (i.e. $C^*=0$), discuss relevant studies subsequent to those by Sparrow \emph{et al.}, including representative ones pertaining to the conjugate problem. Additionally, they provide a procedure using the formulation by~\cite{Sparrow1978} to find the unique combination of fin spacing, thickness and length which minimize the thermal resistance of an LFHS for a prescribed pressure drop driving the flow through it, fin height and fluid-to-solid thermal conductivity ratio. The engineering parameters Po and $\overline{\mathrm{Nu}}$ from~\cite{Sparrow1978} suffice for this. Dense tabulations of them relevant to the optimization of the geometry of LFHSs when the fluid-to-solid thermal conductivity ratio is that of air-to-copper, air-to-aluminum, water-to-copper and water-to-silicon are provided by~\cite{Karamanis-15}.

Subsequent research by~\cite{Karamanis-19b}, again for the fully-shrouded case ($C^*=0$), considered simultaneously-developing flow through an LFHS and relaxed the low Biot number assumption, thereby accounting for temperature variation also across the fin's thickness. Then, the Poiseuille number depends explicitly on the fin thickness-to-height ratio ($t=t^*/H^*$), as well as an additional parameter, i.e., $L^+ = L^*/\left(D_{\mathrm{h}}\mathrm{Re}_{D_\mathrm{h}}\right)$ as per numerical results by~\cite{Curr-72} and, subsequently, many others~\cite{Shah-78}. Thus,~\cite{Karamanis-19b} provide $\overline{\mathrm{Nu}}$ as a function of $\varepsilon$, $t$, $k/k_\mathrm{f}$, $L^+$, $\mathrm{Re}_{D_\mathrm{h}}$ and $\mathrm{Pr}$.
Finally,~\cite{Karamanis-19a} combined the results for Po and $\overline{\mathrm{Nu}}$ from the foregoing studies with flow network modeling and multi-variable optimization to find the optimal fin thickness, spacing, height and length for an array of heat sinks in a circuit pack. 

More recently, the hydrodynamic problem with ($C^*>0$) or without clearance ($C^*=0$) was revisited by~\cite{Miyoshi2024}. They considered the limit of small fin spacing $\varepsilon \ll 1$, which is typical in practice. For example,~\cite{Iyengar-07} report it to range from $1/60$ (bonding) up to $1/6$ (die-casting), or approximately $\varepsilon \approx 0.015 - 0.15$. In this limit,~\cite{Miyoshi2024} presented new flow solutions and $\mathrm{Po}$ formulas for a range of clearances, using complex conformal mapping techniques and matched asymptotic expansions.

The analysis of \cite{Miyoshi2024} was limited to the hydrodynamic problem. In this paper we consider the corresponding conjugate thermal problem (i.e. precisely the one solved by~\cite{Sparrow1978}) in the limit of small fin spacing ($\varepsilon \ll 1$) with finite clearance ($c >0$). We derive explicit formulas for the coupled temperature fields in the fluid and the fin, and the local and overall heat transfer quantities as a function of $\varepsilon$, $c$ and $\Omega$, thereby replacing many of the numerical results of~\cite{Sparrow1978}. To validate these formulas, we compare them to numerical solutions of the full model, as well as the results of Sparrow et al. Two approximations (leading order and higher order, respectively) for the average Nusselt number are found to take the simple forms:
\begin{align}
    \overline{\mathrm{Nu}}^{(0)} &=\frac{2.4304~\varepsilon}{c(2+\varepsilon)}  +O(\varepsilon^2), \\
    \overline{\mathrm{Nu}}^{(1)} &= \frac{\varepsilon}{c(2 + \varepsilon)} \displaystyle{\left[2.4304 - \frac{\varepsilon}{c}\left( 0.5362 + \frac{2.6449}{\Omega}\right)\right]} +O(\varepsilon^3), 
\end{align}
and their comparisons to numerical solutions are summarized in Figs. \ref{fig:Nu_bar}, \ref{fig:error_contours_Nu_bar_1}, \ref{fig:error_contours_Nu_bar_0}. The numerical constants in the above, shown to 4 decimal places, are readily calculated to arbitrary precision.

The paper is structured as follows. The mathematical problem is formulated in section \ref{sec:problem_formulation}. The narrow fin-spacing limit ($\varepsilon \ll 1$) of the flow and thermal problems are presented in section \ref{sec:small-fin-spacing}, with a summary of the theoretical results and formulas in section \ref{sec:temperature_summary}. The Nusselt number definitions and formulas are given in \ref{sec:Nusselt_numbers}, and they are compared to numerical solutions in section \ref{sec:results_comparison}, followed by conclusions in \ref{sec:conclusions}.

\section{Problem formulation}
\label{sec:problem_formulation}
In this section we formulate the problem as given in~\cite{Sparrow1978}. A schematic of the shrouded heat sink is shown in Fig.~\ref{fig:domain_schematic}. The fin array is periodic and aligned longitudinally with the flow direction. The fins are spaced a length $S^*$ apart, with a height $H^*$, and the distance (or clearance)  between their tips and the shroud is $C^*$. The fin thickness is assumed to be negligible compared to other lengths. Throughout, an asterisk will denote that a variable or length is dimensional. We assume that the flow is hydrodynamically- and thermally-fully developed, and consider the coupled (conjugate) thermal problems in the fluid and fins simultaneously.

The flow is unidirectional in the $z^*$ direction. The velocity field, $w^*(x^*,y^*)$, is governed by
\begin{align}
    \frac{\partial^2 w^*}{\partial x^{*2}} + \frac{\partial^2 w^*}{\partial y^{*2}} &= \frac{1}{\mu} \frac{\mathrm{d} p^*}{\mathrm{d} z^*} \quad \text{in } \mathcal{D}, \label{eq:w_eq_dim}
\end{align}
where $\mu$ is the dynamic viscosity and $\mathrm{d} p^* / \mathrm{d} z^*$ is the constant pressure gradient. By periodicity, we restrict attention to a single fin period, $\mathcal{D} = \{0\leq x^* \leq S^*, 0\leq y^* \leq H^* + C^*\}$. The flow satisfies no-slip ($w^*=0$) on the fins ($x^*=0,S^*$ and $0<y^*<H^*$), base ($y^*=0$) and shroud ($y^*=H^*$), and symmetry ($\partial w^*/\partial x^* = 0$) along the centre-line above each fin ($x^*=0,S^*$ and $H^*<y^*<H^* + C^*$).

The thermal energy equation in the fluid takes the form
\begin{align}
    w^* \frac{\partial T^*}{\partial z^*} &= \alpha \left( \frac{\partial^2 T^*}{\partial x^{*2}} + \frac{\partial^2 T^*}{\partial y^{*2}}\right) \quad \text{in } \mathcal{D}, \label{eq:T_eq_dim}
\end{align}
where $T^*(x^*,y^*,z^*)$ is the fluid temperature and $\alpha$ is its thermal diffusivity. We assume that the base is isothermal, at temperature $T_{\mathrm{base}}^*$; hence, the fully developed assumption implies
\begin{align}
    \frac{\partial T^*}{\partial z^*} &= \frac{T^*(x^*,y^*,z^*)-T_{\mathrm{base}}^*}{T_{\mathrm{b}}^*(z^*) - T_{\mathrm{base}}^*}\frac{\mathrm{d} T_{\mathrm{b}}^*}{\mathrm{d} z^*},
\end{align}
where
\begin{align}
    T_{\mathrm{b}}^* &= \frac{\int_{\mathcal{D}} w^*T^*\,\mathrm{d}A^*}{\int_{\mathcal{D}} w^*\,\mathrm{d}A^*},
\end{align}
is the bulk fluid temperature.  Boundary conditions specifying an isothermal base, adiabatic shroud, and symmetry above the fins ($x^*=0$) and between them ($x^*=S^*/2$) are given by
\begin{align}
    T^* &= T_{\mathrm{base}}^* \qquad \text{on } y^*=0, \label{eq:T_BC_1_dim}\\
    \frac{\partial T^*}{\partial y^*} &= 0  \qquad \text{on } y^*=H^* + C^*,\\
    \frac{\partial T^*}{\partial x^*} &= 0  \qquad \text{on } x^* = 0, \quad H^*< y^* < H^*+C^*, \\
    \frac{\partial T^*}{\partial x^*} &= 0  \qquad \text{on } x^* = S^*/2, \quad 0< y^* < H^*+C^*, \label{eq:T_BC_3_dim}
\end{align}
respectively.
On the fin surface, we have continuity of temperature and heat flux with the conduction problem within the fin. The Biot number based on fin thickness is assumed to be small enough that the temperature across its width is approximately constant, with significant temperature variations occurring only along its height\footnote{This can be shown rigorously by considering the limit $t^*/H^*\to 0$ of the 2D conduction problem in the fin, while simultaneously assuming $k_\mathrm{f}/k = O((t^*/H^*)^{-1})$ so that the product $k_\mathrm{f}t^*/(2kH^*) = \Omega $ stays fixed.}. Performing an energy balance across half the width, $t^*$, of the fin (only half is relevant to the domain considered), conduction up the fin is governed by the one-dimensional equation
\begin{align}
    \frac{k_{\mathrm{f}} t^*}{2} \frac{\mathrm{d}^2 T_{\mathrm{f}}^*}{\mathrm{d}y^{*2}} &= -k \left. \frac{\partial T^*}{\partial x^*}\right|_{x^*=0},\qquad \text{for }0<y^*<H^*, \label{eq:T_fin_eq_1_dim}
\end{align}
where $T_{\mathrm{f}}^*$ is the fin temperature, and $k_{\mathrm{f}}$ and $k$ are the thermal conductivities of the fin and fluid, respectively. The sink term on the right hand side corresponds to heat conducting (at each $y^*$ location) out of the fin and into the fluid. There is also temperature continuity between the fin and fluid,
\begin{align}
    T_{\mathrm{f}}^* &= \left. T^*\right|_{x^*=0},\qquad \text{for }0<y^*<H^*. \label{eq:T_continuity_dim}
\end{align}
(Note, axial conduction in $z^*$ in the fin is neglected as per the fully-developed assumption, but $T_\mathrm{f}^*$ does depend on $z^*$ via (\ref{eq:T_continuity_dim}) because $T^*$ depends on $z^*$.) Of course, there is another identical fin at $x^*=S^*$ with the same temperature $T_\mathrm{f}^*$, but in practice we enforce symmetry down the centreline at $x^*=S^*/2$ instead.

Finally, the isothermal condition at the base and adiabatic condition at the tip (due to its negligible surface area) are, respectively,
\begin{align}
    T_{\mathrm{f}}^* &= T_{\mathrm{base}}^*,\quad \text{at } y^*=0,\\
    \frac{\mathrm{d} T_{\mathrm{f}}^*}{\mathrm{d}y^{*}} &= 0,\quad \text{at } y^*=H^*.
    \label{eq:T_fin_eq_4_dim}
\end{align}
\subsection{Nondimensional equations}
The flow problem (\ref{eq:w_eq_dim}) is nondimensionalized by scaling $x^*$ and $y^*$ with $H^*$ and the velocity with $(-\partial p^* / \partial z^*)H^{*2}/\mu$, i.e., introducing,
\begin{align}
    x &= x^*/H^*, &  y &= y^*/H^*, &  w = \frac{\mu }{(-\mathrm{d} p^* / \mathrm{d} z^*)H^{*2}}w^*,
\end{align}
resulting in a nondimensional streamwise momentum 
(Poisson) equation of the form
\begin{align}
    \frac{\partial^2 w}{\partial x^2} + \frac{\partial^2 w}{\partial y^2} &= -1 \quad \text{in } \mathcal{D}. \label{eq:w_eq}
\end{align}
It is subjected to the boundary conditions
\begin{align}
    w &= 0 \qquad \text{on } y=0,\, 1 + c, \\
    w &= 0  \qquad \text{on } x=0,\varepsilon,\quad 0< y < 1,\\
    \frac{\partial w}{\partial x} &= 0  \qquad \text{on } x = 0,\varepsilon, \quad 1< y < 1+c, \label{eq:w_BC_3}
\end{align}
where the nondimensional geometric parameters are then $\varepsilon = S^*/H^*$, the ratio of fin spacing to fin height, and $c = C^*/H^*$, the ratio of clearance to fin height.

As in \cite{Sparrow1978}, we define a nondimensional temperature field in the fluid as
\begin{align}
    T &= \frac{T^*-T_{\mathrm{base}}^*}{T_{\mathrm{b}}^* - T_{\mathrm{base}}^*}, \label{eq:T_dim_scaling}
\end{align}
and a nondimensional streamwise coordinate as
\begin{align}
    z &= \frac{\alpha z^*}{\overline{w}^*H^{*2}}, 
\end{align}
where $\overline{w}^*$ is the average velocity,
\begin{align}
    \overline{w}^* &= \frac{1}{S^*(H^*+C^*)}\int_\mathcal{D} w^* \mathrm{d}A^*.
\end{align}
Expressing (\ref{eq:T_eq_dim}) in nondimensional form, we have 
\begin{align}
     \left( \frac{~w~}{\overline{w}} \right) \lambda T &= \frac{\partial^2 T}{\partial x^2} + \frac{\partial^2 T}{\partial y^2} \quad \text{in } \mathcal{D}, \label{eq:T_eq}
\end{align}
where 
\begin{align}
    \lambda &= \frac{1}{T_{\mathrm{b}}^* - T_{\mathrm{base}}^*}\frac{\mathrm{d} T_b^*}{\mathrm{d} z},
\end{align}
is the nondimensional exponential decay rate of $T_{\mathrm{b}}^* - T_{\mathrm{base}}^*$, which is a negative constant due to the fully-developed assumption. This constant is fixed by enforcing the definition of the bulk temperature, which in nondimensional terms must be equal to one, i.e.,
\begin{align}
    \frac{1}{\varepsilon (1+c)}\int_{\mathcal{D}} \left(\frac{~w~}{\overline{w}}\right)T\, \mathrm{d}A &= 1. \label{eq:lambda_T_eq}
\end{align}
The corresponding boundary conditions (\ref{eq:T_BC_1_dim})-(\ref{eq:T_BC_3_dim}) become
\begin{align}
    T &= 0 \qquad \text{on } y=0, \\
    \frac{\partial T}{\partial y} &= 0  \qquad \text{on } y=1+c,\\
    \frac{\partial T}{\partial x} &= 0  \qquad \text{on } x = 0, \quad 1< y < 1+c,\\
    \frac{\partial T}{\partial x} &= 0  \qquad \text{on } x = \varepsilon/2, \quad 0< y < 1+c.
\end{align}

Defining the nondimensional fin temperature $T_{\mathrm{f}}$ as in (\ref{eq:T_dim_scaling}), where $T$ and $T^*$ are replaced by 
$T_{\mathrm{f}}$ and $T^*_{\mathrm{f}}$, the thermal problem in the fin, (\ref{eq:T_fin_eq_1_dim})-(\ref{eq:T_fin_eq_4_dim}),  becomes
\begin{align}
    \Omega \frac{\mathrm{d}^2 T_{\mathrm{f}}}{\mathrm{d}y^{2}} &= - \left. \frac{\partial T}{\partial x}\right|_{x=0}, \quad \text{for }0<y<1,\label{eq:T_fin_eq_1} \\
    T_{\mathrm{f}} &= \left. T\right|_{x=0}, \quad \text{for }0<y<1, \\
    T_{\mathrm{f}} &= 0,\quad \text{at } y=0,\\
    \frac{\mathrm{d} T_{\mathrm{f}}}{\mathrm{d}y} &= 0,\quad \text{at } y=1, \label{eq:T_fin_eq_4}
\end{align}
where $\Omega$ (given by (\ref{eq:Omega})) is the product of the fin-to-fluid conductivity ratio and the fin's (small) thickness-to-height ratio. The limit $\Omega\rightarrow \infty$ corresponds to an isothermal
fin, where $T=0$ along the entire fin surface.

The above nondimensional system of equations can be solved in their current form, but for consistency with the literature and to yield a more convenient equation for $\lambda$, we instead solve for the scaled temperatures
\begin{align}
    \phi &= \frac{T}{\lambda}, & \phi_{\mathrm{f}} &= \frac{T_{\mathrm{f}}}{\lambda}.
\end{align}
Under this transformation, all equations and boundary conditions remain unchanged ($T$ and $T_{\mathrm{f}}$ replaced with $\phi$ and $\phi_{\mathrm{f}}$, respectively), except for the integral condition (\ref{eq:lambda_T_eq}) which becomes
\begin{align}
   \lambda &= \frac{\varepsilon (1+c)}{\displaystyle{ \int_{\mathcal{D}} \left(\frac{\,w\,}{\overline{w}}\right)\phi\, \mathrm{d}A
 }}, \label{eq:lambda_phi_eq}
\end{align}
with $\lambda$ now appearing explicitly.


\section{Small fin-spacing limit}
\label{sec:small-fin-spacing}
We consider the flow and conjugate thermal problems, i.e., (\ref{eq:w_eq})-(\ref{eq:w_BC_3}) and (\ref{eq:T_eq})-(\ref{eq:lambda_phi_eq}),
respectively, in the geometric limit where the fin spacing is small in comparison to the fin height, i.e., $\varepsilon = S^*/H^* \ll 1$. This is typical of real heat sinks by design, to increase the total surface area for heat transfer; see, e.g., the photographs of heat sinks used in the thermal management of electronics in~\cite{Iyengar-07}. Importantly, we will further assume that the nondimensional tip clearance, $c=C^*/H^*$, will remain of order 1 as we take $\varepsilon \to 0$. This is common for lower power components on a printed wiring board (say, voltage regulators) which use smaller heat sinks than higher power ones (say, a central processing unit). The hydrodynamic problem in this limit has already been considered
in detail by~\cite{Miyoshi2024}, and our current focus is on the subsequent thermal problem, so we only present a brief derivation of the hydrodynamic results.

\subsection{Hydrodynamic problem}

\begin{figure}
    \centering
    \includegraphics[width=1\textwidth]{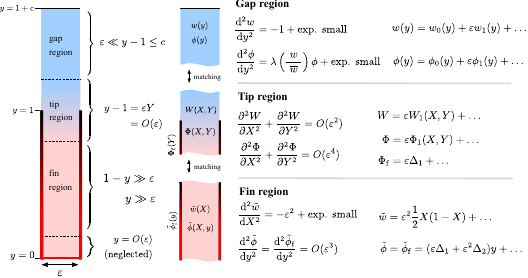}
    \caption{Asymptotic structure of the domain showing the gap, tip and fin regions, and the behaviour of the velocity and temperature expansions in each region (the region close to the base, $y=O(\varepsilon)$, is not considered here). }
    \label{fig:asymptotic_structure}
\end{figure}

It is convenient to rescale the $x$ coordinate by defining $X=x/\varepsilon$, so that the problem domain is independent of $\varepsilon$. Then, (\ref{eq:w_eq})-(\ref{eq:w_BC_3}) becomes
\begin{align}
    \frac{1}{\varepsilon^2}\frac{\partial^2 w}{\partial X^2} + \frac{\partial^2 w}{\partial y^2} &= -1 \quad \text{in } \mathcal{D}=\{0<X<1,\,0<y<1+c\}, \label{eq:w_rescaled_eq}\\
    w &= 0 \qquad \text{on } y=0,\, 1 + c, \label{eq:w_rescaled_BC_1}\\
    w &= 0  \qquad \text{on } X=0,1,\quad 0< y < 1, \label{eq:w_rescaled_BC_2}\\
    \frac{\partial w}{\partial X} &= 0  \qquad \text{on } X = 0,1,\quad 1< y < 1+c. \label{eq:w_rescaled_BC_3}
\end{align}
Considering now $\varepsilon \to 0$, the asymptotic solution takes a different form depending on the region in the domain. Employing matched asymptotic expansions, the domain decomposes into a \emph{gap region} ($1 < y < 1 + c$) above the fins, a \emph{fin region} ($0 < y < 1$) between the fins (but at least a distance $O(\varepsilon)$ away from their base or tips), and a short \emph{tip region} near the fin tips ($y - 1 = O(\varepsilon)$) that transitions between them. There is also a small \emph{base region}, ($y=O(\varepsilon)$) but it is not relevant to the thermal analysis (see \cite{Miyoshi2024} for more details). A schematic of the different regions and the resulting problems within each, is shown in figure \ref{fig:asymptotic_structure}.

\subsubsection*{Gap region: $1 < y \leq 1 + c$}
In the gap region above the fins, a regular expansion $w = w_0 + \varepsilon w_1 + \cdots$ for $\varepsilon \ll1$ leads to the result that $w$ is independent of $X$ (i.e., only a function of $y$) to all algebraic orders. Thus the unit pressure gradient, and the no-slip condition on the shroud ($y=1+c$) lead to a Poiseuille-type parabolic flow profile. The leading order flow is
\begin{align}
    w_0(y) &= -\frac{1}{2}(y-1)(y-1-c), \label{eq:w_0_solution}
\end{align}
and the $O(\varepsilon)$ correction is a shear flow
\begin{align}
    w_1(y) &= -\frac{\log 2}{2\pi}(y - 1 - c), \label{eq:w_1_solution}
\end{align}
that is induced by the non-parabolic flow in the \emph{tip region}, which is discussed after we present the solution in the \emph{fin region}.

\subsubsection*{Fin region: $0 < y < 1$}
Denoting the solution in this region by $\tilde{w}$, a regular expansion $\tilde{w} = \tilde{w}_0 +\varepsilon \tilde{w}_1 + \cdots$ leads this time to a Poiseuille flow but with variation in the $X$ direction. This is because the flow must satisfy no-slip conditions on the fins at $X=0$ and 1. The result is that
\begin{align}
    \tilde{w} &= \frac{1}{2}\varepsilon^2 X(1-X) + \cdots, \label{eq:w_tilde_solution}
\end{align}
where any higher orders are beyond all algebraic powers, and thus are exponentially small in $\varepsilon$.

Interestingly, the flow in both the \emph{gap} and \emph{fin} regions is parabolic, but with variations oriented perpendicularly to one another. The transition between the two solutions takes place in the tip region, where the solution varies in both the $X$
and $y$ directions.

The no-slip condition at the bottom of the domain, $y=0$, can be easily satisfied by the inclusion of a simple series, which is exponentially small unless $y=O(\varepsilon)$, i.e. close to the domain bottom. The modified solution is \cite{Miyoshi2024}
\begin{align}
    \tilde{w}(X,y) &= \frac{1}{2}\varepsilon^2 X(1-X) - 4\varepsilon^2 \sum_{n=1,3,\ldots}\frac{ \mathrm{e}^{-n\pi y/\varepsilon}\sin n\pi X}{n^3\pi^3}, \label{eq:w_tilde_with_base_solution}
\end{align}

This is typically only relevant for visualising the flow field, since the effect of the base region on the average velocity is negligible.

\subsubsection*{Tip region: $y - 1 = O(\varepsilon)$}
Near the fin tips, i.e. an $O(\varepsilon)$ distance away, the variation in $X$ becomes important; therefore, we introduce the inner variable $Y = (y-1)/\varepsilon$ which is $O(1)$ in this region, and we denote the solution here by $w = W(X,Y)$.

It turns out that asymptotic matching with the gap region above ($Y \to +\infty$) implies that the solution is $O(\varepsilon)$ here at leading order,
\begin{align}
    W = \varepsilon W_1(X,Y) + O(\varepsilon^2),
\end{align}
and $W_1(X,Y)$ (from (\ref{eq:w_rescaled_eq}), (\ref{eq:w_rescaled_BC_2}), and (\ref{eq:w_rescaled_BC_3})) satisfies in the infinite strip $\mathcal{D_{\mathrm{tip}}} = \{0<X<1,\,-\infty<Y<\infty\}$, 
\begin{align}
    \nabla_{XY}^2 W_1 &= 0 \qquad \text{in } \mathcal{D_{\mathrm{tip}}}, \label{eq:W_1_eq}\\
    W_1 &= 0  \qquad \text{on } X=0,1,\quad Y < 0, \label{eq:W_1_BC_1}\\
    \frac{\partial W_1}{\partial X} &= 0  \qquad \text{on } X = 0,1,\quad 0< Y, \label{eq:W_1_BC_2}
\end{align}
where $\nabla_{XY}^2 = \partial^2/\partial X^2 + \partial^2/\partial Y^2$. The asymptotic matching conditions with the gap region ($Y\to +\infty$) above and fin region ($Y\to -\infty$) below, are given by
\begin{align}
    W_1 &\sim \frac{1}{2}cY \quad \text{as } Y \to \infty, \label{eq:W_1_matching_up} \\
    W_1 &\to 0 \quad \text{as } Y \to -\infty. \label{eq:W_1_matching_down}
\end{align}
This boundary value problem was solved exactly by \cite{Miyoshi2024} using standard conformal mapping techniques, giving the convenient solution in terms of a complex variable
\begin{align}
    W_1 &= -\frac{c}{2\pi}\log \left|
        \frac{\mathrm{e}^{\mathrm{i}\pi/4} - \tan^{1/2}(\pi Z /2)}{\mathrm{e}^{\mathrm{i}\pi/4} + \tan^{1/2}(\pi Z /2)}
    \right|,\qquad \text{where } Z = X+ \mathrm{i}Y. \label{eq:W_1_solution}
\end{align}
Notably, a constant term that perturbs the shear behaviour as $Y \to \infty$ follows from this solution:
\begin{align}
    W_1 &\sim \frac{1}{2}c\left[
        Y + \frac{\log 2}{\pi} + O(\mathrm{e}^{-\pi Y})
    \right] \quad \text{as } Y \to \infty, \label{eq:W_1_far_field}
\end{align}
and matching at $O(\varepsilon)$ with the gap region solution leads to the response $w_1$, given by (\ref{eq:w_1_solution}).

\subsubsection*{Composite flow solutions}
A pair of composite flow solutions valid through larger regions of the domain can be constructed by adding solutions in adjacent regions and subtracting the solution in the overlap region between them. When patched together in a piece-wise fashion, these can be used to construct a single global flow field if desired. We do so by splitting the domain into $y\geq 1$ and $y<1$ (i.e., at the fin tips), then constructing a composite in each region separately.

A gap--tip composite (restricted to $1 \leq y \leq 1+c$) is given by 
\begin{align}
    w_{\mathrm{gap-tip}} &= \underbrace{w_0(y) + \varepsilon w_1(y)}_{\text{gap region}} + \underbrace{\varepsilon W_1 \left(X,\frac{y-1}{\varepsilon}\right)}_{\text{tip region}} - \underbrace{\frac{ c}{2}\left(y - 1 + \frac{\varepsilon \log 2}{\pi}\right)}_{\text{gap--tip overlap}}, \nonumber\\
    &= -\frac{1}{2}(y-1-c)\left(y-1 + \frac{\varepsilon \log 2}{\pi}\right) + \varepsilon W_1 \left(X,\frac{y-1}{\varepsilon}\right),\qquad 0<X<1. \label{eq:composite_velocity_upper}
\end{align}

On the other hand, a one-term composite (say restricted to $0 \leq y < 1$) between the tip and fin regions is given simply by the tip solution $\varepsilon W_1$, since it matches with ``zero" appearing at that order in the fin region. Strictly, to bring in the $O(\varepsilon^2)$ parabolic flow from the fin region one would require a two-term composite, but we do not possess a second term (of $O(\varepsilon^2)$) in the tip region. However, we can form an \emph{ad hoc} composite, with an $O(\varepsilon^2)$ error in the tip region, by simply superimposing the tip and fin region solutions:
\begin{align}
    w_{\mathrm{tip-fin}} &=  \underbrace{\varepsilon W_1 \left(X,\frac{y-1}{\varepsilon}\right)}_{\text{tip region}} + \underbrace{\tilde{w}(X,y)}_{\text{fin region}},\qquad 0<X<1. \label{eq:composite_velocity_lower}
\end{align}
Here $W_1(X,Y)$ is (\ref{eq:W_1_solution}) and $\tilde{w}(X,y)$ is (\ref{eq:w_tilde_with_base_solution}).
Then, the correct leading order behaviour is exhibited in each of the (tip and fin) regions.

\subsubsection*{Poiseuille number}
The mean velocity is given by
\begin{align}
    \overline{w} &= \frac{1}{1+c}\int_0^{1} \int_0^{1+c} w\,\mathrm{d}y\mathrm{d}X = \frac{c^3}{12(1+c)}\left[
        1 + \frac{\varepsilon \log 8}{\pi c}
    \right]+ O(\varepsilon^2),  \label{eq:average_w_solution}
\end{align}
where the contributions from the tip region  (velocity of $O(\varepsilon)$ over a region of area $O(\varepsilon)$) and fin region (velocity of $O(\varepsilon^2)$ over a region of area $O(1)$) are both $O(\varepsilon^2)$, so $\overline{w}$ up to $O(\varepsilon)$ follows only from the gap solution (\ref{eq:w_0_solution}), (\ref{eq:w_1_solution}).

The friction factor is defined as \cite{Sparrow1978}
\begin{align}
    f &= \frac{(-\mathrm{d}p^*/\mathrm{d}z^*) D_e^*}{\frac{1}{2}\rho \overline{w}^{*2}}, & D_e^* &= \frac{4(H^* + C^*)S^*}{2(H^* + S^*)}, \label{eq:f_definition} 
\end{align}
and the Poiseuille number is then $\mathrm{Po}=f\mathrm{Re}$ where $\mathrm{Re}=\rho \overline{w}^*D_e^*/\mu$ is the Reynolds number. In terms of nondimensional quantities,
\begin{align}
    \mathrm{Po} = f\mathrm{Re} &= \frac{8(1+c)^2\varepsilon^2}{\overline{w}(1+\varepsilon)^2}. \label{eq:fRe_definition}
\end{align}
 Substituting the asymptotic solution (\ref{eq:average_w_solution}) for $\overline{w}$ results in the elementary expression \cite{Miyoshi2024}
\begin{align}
    \mathrm{Po} &= \frac{96(1+c)^3\varepsilon^2}{\displaystyle{
        c^2\left(c + \frac{\varepsilon \log 8}{\pi}\right)(1+\varepsilon)^2
    }} + O(\varepsilon^4),\quad \text{as }\varepsilon \to 0. \label{eq:fRe_solution}
\end{align}
%
This expression was compared to an exact solution valid for arbitrary $\varepsilon$ in \cite{Miyoshi2024}, and shown to be accurate to within 15\% if $\varepsilon \lesssim 0.3 c$. As expected, the approximation breaks down when $c$ becomes small (i.e. comparable to $\varepsilon$) thus, for a given accuracy, the range of $\varepsilon$ must shrink to maintain validity.

\subsection{Thermal problem}

In the narrow-fin-spacing limit, the solution to the thermal problem (\ref{eq:T_eq})-(\ref{eq:lambda_phi_eq}) will depend on the asymptotic region considered, just as for the hydrodynamic problem. With the hydrodynamic solution summarized, we now proceed with the analysis of the thermal problem. Again, the different regions and the thermal problems in each, are summarized in figure \ref{fig:asymptotic_structure}. 

Rescaling $x = X/\varepsilon$ and defining $\mathcal{W} = w/\overline{w}$ which is now (approximately) known, the problem for $\phi$ is
\begin{align}
    \lambda \mathcal{W} \phi &= \frac{1}{\varepsilon^2}\frac{\partial^2 \phi}{\partial X^2} + \frac{\partial^2 \phi}{\partial y^2} \quad \text{in } \mathcal{D}, \label{eq:phi_scaled_eq} \\
    \phi &= 0 \qquad \text{on } y=0, \\
    \frac{\partial \phi}{\partial y} &= 0  \qquad \text{on } y=1+c,\\
    \frac{\partial \phi}{\partial X} &= 0  \qquad \text{on } X = 0, \quad 1< y < 1+c, \\
    \frac{\partial \phi}{\partial X} &= 0  \qquad \text{on } X = 1/2, \quad 0< y < 1+c. \label{eq:phi_scaled_eq_5}
\end{align}
in the fluid, and
\begin{align}
    \varepsilon\Omega \frac{\mathrm{d}^2 \phi_{\mathrm{f}}}{\mathrm{d}y^{2}} &= - \left. \frac{\partial \phi}{\partial X}\right|_{X=0}, \quad \text{for }0<y<1,\label{eq:phi_fin_scaled_eq_1} \\
    \phi_{\mathrm{f}} &= \left. \phi\right|_{X=0}, \quad \text{for }0<y<1, \\
    \phi_{\mathrm{f}} &= 0,\quad \text{at } y=0,\\
    \frac{\mathrm{d} \phi_{\mathrm{f}}}{\mathrm{d}y} &= 0,\quad \text{at } y=1. \label{eq:phi_fin_scaled_eq_4}
\end{align}
in the fin, with decay constant
\begin{align}
   \lambda &= \frac{1+c}{\displaystyle{ \int_0^1 \int_0^{1+c} \mathcal{W}\phi\, \mathrm{d}y\mathrm{d}X
 }}. \label{eq:lambda_phi_scaled_eq}
\end{align}

\subsubsection*{Gap region: $1<y\leq 1+c$}

Given $\phi = 0$ on the base, we expect $\phi$ to have the largest magnitude in the gap region, furthest from the heat transport surfaces. If we anticipate a balance between conduction in the $y$-direction and streamwise advection, then $\partial^2 \phi/\partial y^2 \sim \lambda \mathcal{W}\phi$. But $\mathcal{W}=O(1)$ in the gap region and $\lambda = O(1/\phi)$ from (\ref{eq:lambda_phi_scaled_eq}), and thus $\partial^2 \phi/\partial y^2  = O(1)$. This gives the leading order estimate that $\phi = O(1)$ and $\lambda = O(1)$ as $\varepsilon \to 0$. Considering expansions 
\begin{align}
    \phi &= \phi_0 + \varepsilon \phi_1 + O(\varepsilon^2),\\
    \lambda &= \lambda_0 + \varepsilon \lambda_1 + O(\varepsilon^2),\\
    \mathcal{W} &= \mathcal{W}_0(y) + \varepsilon \mathcal{W}_1(y) + O(\varepsilon^2),
\end{align}
similar arguments to those for the velocity field in the gap region follow here and we find that $\phi_0(y)$, $\phi_1(y)$ depend on $y$ only. At leading order, $\phi_0(y)$ satisfies the one-dimensional problem
\begin{align}
   \lambda_0 \mathcal{W}_0(y) \phi_0 &= \frac{\mathrm{d}^2 \phi_0}{\mathrm{d} y^2} \quad \text{in } 1<y<1+c, \label{eq:phi_0_eq} \\
    \frac{\mathrm{d}\phi_0}{\mathrm{d} y} &= 0  \qquad \text{on } y=1+c, \label{eq:phi_0_BC_1}\\ 
    \lambda_0 &= \frac{1+c}{\displaystyle{ \int_1^{1+c} \mathcal{W}_0(y)\phi_0(y)\, \mathrm{d}y
    }}, \label{eq:lambda_phi_0_eq}
\end{align}
and at first order, $\phi_1(y)$ satisfies
\begin{align}
   (\lambda_1 \mathcal{W}_0  + \lambda_0 \mathcal{W}_1) \phi_0 + \lambda_0 \mathcal{W}_0 \phi_1 &= \frac{\mathrm{d}^2 \phi_1}{\mathrm{d} y^2} \quad \text{in } 1<y<1+c, \label{eq:phi_1_eq} \\
    \frac{\mathrm{d}\phi_1}{\mathrm{d} y} &= 0  \qquad \text{on } y=1+c,\\ 
    \lambda_1 &= -\frac{\lambda_0^2}{1+c} \displaystyle{ \int_1^{1+c} (\mathcal{W}_0\phi_1 + \mathcal{W}_1\phi_0)\, \mathrm{d}y
    }. \label{eq:lambda_phi_1_eq}   
\end{align}
Note, since the velocity $\mathcal{W}=O(\varepsilon)$ and $O(\varepsilon^2)$ in the tip ($y-1=O(\varepsilon)$) and fin regions ($0<y<1$), the contributions of those regions to the integral for $\lambda$ appear at $O(\varepsilon^2)$ and hence are ignored.

These problems for $\phi_0$ and $\phi_1$ also have a matching condition with the tip region as $y \to 1^+$, which are yet to be determined. Nonetheless, we will solve for $\phi_0$ and $\phi_1$ numerically, detailed later in the summary, section \ref{sec:temperature_summary}.

\subsubsection*{Fin region: $0\leq y<1$}
In the fin region,  we must solve for the fluid temperature but also the fin temperature $\phi_{\mathrm{f}}(y)$. Denoting the fluid temperature here by $\tilde{\phi}$, and expanding both temperatures as
\begin{align}
    \tilde{\phi} &= \tilde{\phi}_0 + \varepsilon \tilde{\phi}_1 + O(\varepsilon^2),\\
    \phi_{\mathrm{f}} &= \phi_{\mathrm{f}0} + \varepsilon \phi_{\mathrm{f}1} + O(\varepsilon^2),
\end{align}
and noting that the advection terms are $O(\varepsilon^2)$ at most (since $\mathcal{W} = O(\varepsilon^2)$ here), at leading order in (\ref{eq:phi_scaled_eq}), (\ref{eq:phi_scaled_eq_5}) and (\ref{eq:phi_fin_scaled_eq_1}) we find
\begin{align}
    \frac{\partial^2 \tilde{\phi}_0}{\partial X^2} &= 0,\\
    \frac{\partial \tilde{\phi}_{0}}{\partial X} &= 0, \quad \text{at }X=0,1/2
\end{align}
and so integrating the former and applying the latter gives that $\tilde{\phi}_0(y)$ is a function of $y$ only, and matches the fin temperature by continuity, $\phi_{\mathrm{f}0}(y) = \tilde{\phi}_0(y)$. At the next order, $O(\varepsilon^{-1})$ in (\ref{eq:phi_scaled_eq}) and $O(\varepsilon)$ in (\ref{eq:phi_fin_scaled_eq_1}), we have
\begin{align}
    \frac{\partial^2 \tilde{\phi}_1}{\partial X^2} &= 0, \label{eq:phi_tilde_1_eq}\\
    \Omega \frac{\mathrm{d}^2 \phi_{\mathrm{f}0}}{\mathrm{d}y^{2}} &= - \left. \frac{\partial \tilde{\phi_1}}{\partial X}\right|_{X=0}, \quad \text{for }0<y<1. \label{eq:phi_fin_0_eq}
\end{align}
Integrating (\ref{eq:phi_tilde_1_eq}), and applying symmetry about the midline $X=1/2$ then $\tilde{\phi}_1(y)$ is also only a function of $y$, with $\phi_{\mathrm{f}1}(y) = \tilde{\phi}_1(y)$ by continuity. Then the right hand side of (\ref{eq:phi_fin_0_eq}) vanishes; no heat is conducting out of the fin at this order in this region. Integrating (\ref{eq:phi_fin_0_eq}) and applying the isothermal condition $\phi_{\mathrm{f}0}=0$ at the base $y=0$, we get
\begin{align}
    \phi_{\mathrm{f}0}(y) &= \tilde{\phi}_0(y) = \Delta_0 y,
\end{align}
where $\Delta_0$ is a constant to be determined. Going to the next order, $O(\varepsilon^0)$ in (\ref{eq:phi_scaled_eq}) and $O(\varepsilon^2)$ in (\ref{eq:phi_fin_scaled_eq_1}), similar arguments give that
\begin{align}
    \phi_{\mathrm{f}1}(y) &= \tilde{\phi}_1(y) = \Delta_1 y,
\end{align}
for a constant $\Delta_1$. We cannot apply the boundary condition (\ref{eq:phi_fin_scaled_eq_4}) at the ridge tip $y=1$ at each order since the assumption that the fluid temperature ($\tilde{\phi}_0$ and $\tilde{\phi}_1$) is independent of $X$ breaks down there. Hence, we must match with a solution in the tip region. Nonetheless, it is remarkable that there is no conduction out of (the majority of) the fin into the fluid at the first two orders of the expansion; conduction is purely one-dimensional and in the $y$ direction.

\subsubsection*{Tip region: $y-1 = O(\varepsilon)$}
Estimates from the gap and fin region solutions as $y \to 1$ are that the temperatures, denoted here by $\phi=\Phi$ and $\phi_{\mathrm{f}} = \Phi_{\mathrm{f}}$, are $O(1)$. Substituting $y = 1 + \varepsilon Y$ into (\ref{eq:phi_scaled_eq})-(\ref{eq:phi_fin_scaled_eq_4}), and noting that $\mathcal{W}\sim \varepsilon W_1 / \overline{w} = O(\varepsilon)$ here, we find
\begin{align}
    \nabla_{XY}^2\Phi &= O(\varepsilon^3)\quad \text{in } \mathcal{D}_{\mathrm{tip}}=\{0<X<1/2,\,-\infty<Y<\infty\}, \label{eq:Phi_eq} \\
    \frac{\partial \Phi}{\partial X} &= 0  \qquad \text{on } X = 0, \quad 0< Y, \label{eq:Phi_BC_1} \\
    \frac{\partial \Phi}{\partial X} &= 0  \qquad \text{on } X = 1/2, \quad -\infty < Y < \infty, \label{eq:Phi_BC_2}
\end{align}
in the fluid, and
\begin{align}
    \Omega \frac{\mathrm{d}^2 \Phi_{\mathrm{f}}}{\mathrm{d}Y^{2}} &= - \varepsilon\left. \frac{\partial \Phi}{\partial X}\right|_{X=0}, \quad \text{for }Y<0, \label{eq:Phi_f_eq}\\
    \Phi_{\mathrm{f}} &= \left. \Phi\right|_{X=0}, \quad \text{for }Y<0, \label{eq:Phi_f_contin}\\
    \frac{\mathrm{d} \Phi_{\mathrm{f}}}{\mathrm{d}Y} &= 0,\quad \text{at } Y=0, \label{eq:Phi_f_BC}
\end{align}
in the fin. Expanding as per
\begin{align}
    \Phi &= \Phi_0 + \varepsilon \Phi_1 + O(\varepsilon^2), \\
    \Phi_{\mathrm{f}} &= \Phi_{\mathrm{f}0} + \varepsilon \Phi_{\mathrm{f}1} + O(\varepsilon^2),
\end{align}
then $\Phi_{\mathrm{f}0}$ satisfies $\mathrm{d}^2 \Phi_{\mathrm{f}0} / \mathrm{d}Y^2 = 0$, which integrated with the application of boundary condition (\ref{eq:Phi_f_BC}) implies that $\Phi_{\mathrm{f}0}$ is a constant. Matching (as $Y\to -\infty$) with the fin region solution (as $y\to 1^-$) at leading order means
\begin{align}
    \Phi_{\mathrm{f}0} &\to \Delta_0, \quad \text{as }Y \to -\infty,
\end{align}
and hence $\Phi_{\mathrm{f}0}\equiv \Delta_0$. With the fin isothermal, the fluid temperature $\Phi_0$ satisfies $\nabla_{XY}^2\Phi_0 = 0$ in the strip, with $\Phi_0 = \Delta_0$ on the fin ($Y<0$), symmetry above the fins ($Y>0$), and matching conditions
\begin{align}
    \Phi_0 &\to \phi_0(y=1^+), \quad \text{as }Y \to \infty, \\
    \Phi_0 &\to \Delta_0, \quad \text{as }Y \to -\infty,
\end{align}
with the gap and fin regions, respectively. However, the problem for $\Phi_0 - \Delta_0$ is identical in form to the velocity problem $W_0$ in this region, which was found to be identically zero \cite{Miyoshi2024}. Therefore, $\Phi_0 \equiv \Delta_0$ and $\phi_0(y=1^+) = \Delta_0$, with $\Delta_0$ still not known.

Proceeding to the next order, the fin temperature correction $\Phi_{\mathrm{f}1}$ satisfies $\mathrm{d}^2 \Phi_{\mathrm{f}1} / \mathrm{d}Y^2 = 0$. Integrating and applying (\ref{eq:Phi_f_BC}) implies that $\Phi_{\mathrm{f}1}$ is also a constant. The matching condition with the fin region solution at $O(\varepsilon)$ means
\begin{align}
    \Phi_{\mathrm{f}1} &\sim \Delta_0 Y + \Delta_1 \quad \text{as }Y \to -\infty,
\end{align}
but as $\Phi_{\mathrm{f}1}$ is constant, this is possible only if $\Delta_0 = 0$, giving $\Phi_{\mathrm{f}1}\equiv \Delta_1$. The leading order fin and fluid temperatures are thus actually $O(\varepsilon)$. Also, since $\Delta_0=0$, the matching condition for the gap solution is now 
\begin{align}
    \phi_0 &= 0 \quad \text{at }y=1^+. \label{eq:phi_0_matching_down}
\end{align}

The problem for $\Phi_1$ follows from (\ref{eq:Phi_eq})-(\ref{eq:Phi_BC_2}), (\ref{eq:Phi_f_contin}) as
\begin{align}
    \nabla_{XY}^2\Phi_1 &= 0\quad \text{in } \mathcal{D}_{\mathrm{tip}}, \label{eq:Phi_1_eq}\\
    \frac{\partial \Phi_1}{\partial X} &= 0  \qquad \text{on } X = 0,1, \quad 0< Y, \\
    \Phi_1 &= \Delta_1  \qquad \text{on } X = 0,1, \quad Y < 0. 
\end{align}
with matching condition to the fin region,
\begin{align}
    \Phi_{1} &\to \Delta_1 \quad \text{as }Y \to -\infty. \label{eq:Phi_1_matching_down}
\end{align}
For the matching condition (as $Y \to +\infty$) with the gap region, it is convenient to derive one for the heat flux. Integrating the leading order gap equation (\ref{eq:phi_0_eq}) across the gap $1<y<1+c$ and using (\ref{eq:phi_0_BC_1}), (\ref{eq:lambda_phi_0_eq}), we find
\begin{align}
    \frac{\mathrm{d}\phi_0}{\mathrm{d}y} &= -(1+c),\quad \text{at }y=1^+. \label{eq:phi_0_heat_flux}
\end{align}
Doing the same for the first order equation (\ref{eq:phi_1_eq}), we find
\begin{align}
    \frac{\mathrm{d}\phi_1}{\mathrm{d}y} &= 0,\quad \text{at }y=1^+.
\end{align}
Transforming to the tip variable $Y = (y-1)/\varepsilon$, this gives the matching condition on $\Phi_1$,
\begin{align}
    \frac{\partial \Phi_1}{\partial Y} &\to -(1+c) \quad \text{as } Y \to \infty, \\
    \text{or }\Phi_1 &\sim -(1+c)Y  \quad \text{as } Y \to \infty. \label{eq:Phi_1_matching_up}
\end{align}
Notice now that the problem (\ref{eq:Phi_1_eq})-(\ref{eq:Phi_1_matching_down}), (\ref{eq:Phi_1_matching_up}), when considered as a problem for the difference $\Phi_1 - \Delta_1$, is identical to the problem for $W_1$ (see (\ref{eq:W_1_eq})-(\ref{eq:W_1_matching_down})), but with a ``shear-rate" of $-(1+c)$ as $Y\to \infty$ instead of $c/2$. Therefore, remarkably, the same solution can be used here for the thermal problem, and it is given by
\begin{align}
    \Phi_1 &= \Delta_1  - \frac{2(1+c)}{c}W_1(X,Y),\\
        &= \Delta_1  + \frac{1+c}{\pi}\log \left|
        \frac{\mathrm{e}^{\mathrm{i}\pi/4} - \tan^{1/2}(\pi Z /2)}{\mathrm{e}^{\mathrm{i}\pi/4} + \tan^{1/2}(\pi Z /2)}
    \right|,\qquad \text{where } Z = X+ \mathrm{i}Y, \label{eq:Phi_1_solution}
\end{align}
with far field behaviour
\begin{align}
    \Phi_1 &\sim -(1+c)\left[
        Y + \frac{\log 2}{\pi} - \frac{\Delta_1}{1+c} + O(\mathrm{e}^{-\pi Y})
    \right] \quad \text{as } Y \to \infty. \label{eq:Phi_1_far_field}
\end{align}
We have yet to determine the constant $\Delta_1$. To do so, we must proceed to $O(\varepsilon^2)$ in the fin (\ref{eq:Phi_f_eq}), where
\begin{align}
    \Omega \frac{\mathrm{d}^2 \Phi_{\mathrm{f}2}}{\mathrm{d}Y^{2}} &= - \left. \frac{\partial \Phi_1}{\partial X}\right|_{X=0}, \quad \text{for }Y<0. \label{eq:Phi_f_2_eq}
\end{align}
The flux on the right hand side can be evaluated exactly (see Appendix \ref{sec:heat_flux_derivation}) given the solution (\ref{eq:Phi_1_solution}) to give
\begin{align}
    - \left. \frac{\partial \Phi_1}{\partial X}\right|_{X=0} & = \frac{1+c}{\sqrt{\mathrm{e}^{-2\pi Y}-1}}, \quad \text{for }Y<0. \label{eq:Phi_1_surface_flux}
\end{align}
Substituting, and integrating (\ref{eq:Phi_f_2_eq}) from $Y=0$ to $Y<0$ applying (\ref{eq:Phi_f_BC}), we find the heat flux within the fin
\begin{align}
    -\Omega \frac{\mathrm{d} \Phi_{\mathrm{f}2}}{\mathrm{d}Y} &= \frac{1+c}{\pi}\tan ^{-1}\left( \sqrt{\mathrm{e}^{-2\pi Y}-1} \right), \quad \text{for }Y<0. \label{eq:Phi_f_2_solution}
\end{align}
In the limit $Y \to -\infty$, this gives
\begin{align}
    -\Omega \frac{\mathrm{d} \Phi_{\mathrm{f}2}}{\mathrm{d}Y} &\to \frac{1+c}{2}, \quad \text{as }Y\to -\infty, \label{eq:Phi_f_2_matching_down}
\end{align}
which states that the heat flux ($(1+c)/2$) entering the tip region via conduction up a fin is equal to the heat flux leaving (one half of) the tip region via the fluid ($(1+c)/2$). This is because all the heat transfer between the fin and fluid occurs in the tip region. Recall that the solution in the fin region is $\tilde{\phi}=\tilde{\phi}_\mathrm{f}=\varepsilon \Delta_1 y + \cdots$, and substituting $y=1+\varepsilon Y$, the heat flux along the fins must match with (\ref{eq:Phi_f_2_matching_down}) at $O(\varepsilon^2)$, implying that
\begin{align}
    \Delta_1 &= -\frac{1+c}{2\Omega}.
\end{align}
With $\Delta_1$ determined, the final necessary condition on $\phi_1$ in the gap region comes from matching with the constant term in (\ref{eq:Phi_1_far_field}), giving
\begin{align}
    \phi_1 &= -(1+c)\left(\frac{\log 2}{\pi} + \frac{1}{2\Omega}\right)\quad \text{at }y=1^+, \label{eq:phi_1_matching_down}
\end{align}
which closes the problem for $\phi_1$---we discuss its solution in section \ref{sec:temperature_summary}. This finishes the determination of the temperature field to $O(\varepsilon)$ throughout the whole domain.

\subsubsection*{Second order in the fin region}
The heat flux leaving the tip can also be determined at the next order, $O(\varepsilon^2)$, and so can the corresponding temperature correction throughout the entire fin. Although $\Phi_2$ throughout the tip region is difficult to find, from a global energy in that region one can determine the total heat flux leaving the fin, $\int_{-\infty}^0 -\partial \Phi_2/\partial X~\mathrm{d}Y$. Then the fin equation (\ref{eq:Phi_f_eq}) at third order can be integrated and matched with the correction in the fin region. The analysis is given in Appendix \ref{sec:second_order_tip_fin_regions}, and the $O(\varepsilon^2)$ term in the fin region ($0\leq y <1$, $1-y \gg \varepsilon)$ is
\begin{align}
    \tilde{\phi}_2(y) &= \phi_{\mathrm{f}2}(y) = \frac{1+c}{4\Omega^2}y,
\end{align}
which is linear in $y$, similar to leading order.

\section{Summary of temperature solution}
\label{sec:temperature_summary}
With the asymptotic structure and solution for the temperature field derived in section \ref{sec:small-fin-spacing} up to $O(\varepsilon)$ (and $\varepsilon^2$ in the fin region), we now provide a summary of the thermal results. Also, see figure \ref{fig:asymptotic_structure} for a visual summary.

\subsubsection*{Fin region ($0\leq y<1$, $1-y \gg \varepsilon$)}
Here the temperature field is isothermal in the $X$ direction, with only linear variation in the $y$ direction corresponding to conduction up the fins:
\begin{align}
   \phi_{\mathrm{f}} &= \tilde{\phi} = \varepsilon \tilde{\phi}_1 + \varepsilon^2 \tilde{\phi}_2 + O(\varepsilon^3) = - (1+c)\left(\frac{\varepsilon}{2\Omega} - \frac{\varepsilon^2}{4\Omega^2}\right) y + O(\varepsilon^3). \label{eq:phi_tilde_summary}
\end{align}
In the case where the fins have infinite conductivity $\Omega = \infty$ (i.e. are isothermal), then $\tilde{\phi}\equiv \phi_{\mathrm{f}} = 0$ to the order considered. 

\subsubsection*{Tip region ($y -1 = \varepsilon Y$, $Y=O(1)$)}
Here the temperature field becomes two-dimensional but the problem is still pure conduction with advection negligible. The temperature in the fluid is
\begin{align}
    \phi &= \varepsilon \Phi_1(X,Y) + O(\varepsilon^2) \quad \text{in } \mathcal{D}_{\mathrm{tip}}=\{0<X<1,\,-\infty<Y<\infty\},
\end{align}
where $\Phi_1$ is given in closed-form by (\ref{eq:Phi_1_solution}), with far-field behaviours
\begin{align}
    \Phi_1 &\sim -(1+c)\left[
        Y + \frac{\log 2}{\pi} + \frac{1}{2\Omega} + O(\mathrm{e}^{-2\pi Y})
    \right] \quad \text{as } Y \to \infty, \\    
    \Phi_1 &\to -\frac{1+c}{2\Omega} \quad \text{as } Y \to -\infty. 
\end{align}
The fin temperature near the tips is simply constant at leading order,
\begin{align}
    \phi_{\mathrm{f}} &= \varepsilon \Phi_{\mathrm{f}1} + O(\varepsilon^2) = -\frac{\varepsilon(1+c)}{2\Omega} + O(\varepsilon^2)
\end{align}
and the $O(\varepsilon^2)$ correction is found by integrating (\ref{eq:Phi_f_2_solution}), giving (\ref{eq:Phi_f_2_full_solution}).

\subsubsection*{Gap region ($1<y\leq 1+c$, $y-1 \gg \varepsilon$)}
Finally, in this region above the fins, the problem up to $O(\varepsilon)$ is independent of $X$, i.e. $\phi = \phi_0(y) + \varepsilon \phi_1(y) + O(\varepsilon^2)$, $\lambda = \lambda_0 + \varepsilon \lambda_1 + O(\varepsilon^2)$. The leading order problem for ($\phi_0, \lambda_0$) is the classical problem with an isothermal, no-slip ``lower wall" at $y=1$ and an adiabatic, no-slip upper wall at $y=1+c$. This problem, (\ref{eq:phi_0_eq})-(\ref{eq:lambda_phi_0_eq}) and (\ref{eq:phi_0_matching_down}), can be transformed to the usual scaling, independent of the clearance $c$, if we define
\begin{align}
    \hat{y} &= (y-1)/c, & \widehat{\mathcal{W}}_0(\hat{y}) &= \mathcal{W}_0(y)c/(1+c), \\
    \hat{\phi}_0 &= \phi_0 / [c(1+c)], & \hat{\lambda}_0 &= c(1+c)\lambda_0,
\end{align}
resulting in
\begin{align}
   \hat{\lambda}_0 \widehat{\mathcal{W}}_0(\hat{y}) \hat{\phi}_0 &= \frac{\mathrm{d}^2 \hat{\phi}_0}{\mathrm{d} \hat{y}^2} \quad \text{in } 0<\hat{y}<1, \label{eq:phi_0_hat_eq} \\
    \frac{\mathrm{d}\hat{\phi}_0}{\mathrm{d} \hat{y}} &= 0  \qquad \text{on } \hat{y}=1, \label{eq:phi_0_hat_BC_1}\\ 
    \hat{\phi}_0 &= 0  \qquad \text{on } \hat{y}=0, \label{eq:phi_0_hat_BC_2}\\ 
    \hat{\lambda}_0 &= \frac{1}{\displaystyle{ \int_0^{1} \widehat{\mathcal{W}}_0(\hat{y})\hat{\phi}_0(\hat{y})\, \mathrm{d}\hat{y}
    }}, \label{eq:lambda_phi_0_hat_eq}
\end{align}
with normalised velocity $\widehat{\mathcal{W}}_0(\hat{y}) = 6\hat{y}(1-\hat{y})$. This problem for $(\hat{\phi}_0,\hat{\lambda}_0)$ has no parameters and only needs to be solved once. It is easily solved numerically, and we do so using Chebyshev collocation methods to discretize space \cite{TrefethenSpectralMethods} and employing an iterative procedure similar to that of \cite{Sparrow1978,Karamanis-16}. That is, given a guess for $\hat{\phi}_0$, we compute $\hat{\lambda}_0$ and the left hand side of (\ref{eq:phi_0_hat_eq}). Given this left hand side, (\ref{eq:phi_0_hat_eq}) is solved together with conditions (\ref{eq:phi_0_hat_BC_1})-(\ref{eq:phi_0_hat_BC_2}), giving a new estimate for $\hat{\phi}_0$. The process is repeated, and grid points increased, until $\hat{\lambda}_0$ changes by less than $10^{-5}$. We arrive at the value (to 4 decimal places) 
\begin{align}
    \hat{\lambda}_0 & \approx -2.4304
\end{align}
The first order problem for ($\phi_1, \lambda_1$), given by (\ref{eq:phi_1_eq})-(\ref{eq:lambda_phi_1_eq}) and (\ref{eq:phi_1_matching_down}), may be transformed in a similar way,
\begin{align}
    \hat{y} &= (y-1)/c, & \widehat{\mathcal{W}}_1(\hat{y}) &= \mathcal{W}_1(y)c^2/(1+c), \\
    \hat{\phi}_1 &= \phi_1 / (1+c), & \hat{\lambda}_1 &= c^2(1+c)\lambda_1,
\end{align}
resulting in
\begin{align}
   (\hat{\lambda}_1 \widehat{\mathcal{W}}_0  + \hat{\lambda}_0 \widehat{\mathcal{W}}_1) \hat{\phi}_0 + \hat{\lambda}_0 \widehat{\mathcal{W}}_0 \hat{\phi}_1 &= \frac{\mathrm{d}^2 \hat{\phi}_1}{\mathrm{d} \hat{y}^2} \quad \text{in } 0<\hat{y}<1, \label{eq:phi_1_hat_eq}\\
    \frac{\mathrm{d}\hat{\phi}_1}{\mathrm{d} \hat{y}} &= 0  \qquad \text{on } \hat{y}=1,\label{eq:phi_1_hat_BC_1}\\
    \hat{\phi}_1 &= -\left(\frac{\log 2}{\pi} + \frac{1}{2\Omega}\right)  \qquad \text{on } \hat{y}=0,\label{eq:phi_1_hat_BC_2}\\ 
    \hat{\lambda}_1 &= -\hat{\lambda}_0^2 \displaystyle{ \int_0^{1} (\widehat{\mathcal{W}}_0\hat{\phi}_1 + \widehat{\mathcal{W}}_1\hat{\phi}_0)\, \mathrm{d}\hat{y}
    },   \label{eq:lambda_phi_1_hat_eq}
\end{align}
where $\widehat{\mathcal{W}}_1 = 6(1-\hat{y})(1-3\hat{y})\log 2/\pi$. This problem only depends on the fin conductivity $\Omega$, but since it is linear, it can be shown (Appendix \ref{sec:solution_for_lambda_hat_1}) that the solution depends linearly on $1/\Omega$, with
\begin{align}
    \hat{\lambda}_1 &= b_0 + \frac{b_1}{2\Omega}
\end{align}
where $b_0,b_1$ are numerical constants. Using similar numerical methods to those we employed to find $\hat{\lambda}_0$, although even simpler since no iteration is needed, we find that they evaluate to (4 decimal places):
\begin{align}
    b_0 &\approx 0.5362, & b_1 & \approx 5.2898. \label{eq:b_0_and_b_1_values}
\end{align}
Consequently, the solution for $\lambda$ and its dependence on all the parameters ($\varepsilon$, $c$ and $\Omega$) is completely determined. It will be useful to consider two levels of approximation, given by
\begin{align}
    \lambda^{(0)} &= \lambda_0 = \frac{\hat{\lambda}_0}{c(1+c)}, & \text{(one term)} \label{eq:lambda_leading_order}\\
    \lambda^{(1)} &= \lambda_0 + \varepsilon \lambda_1 = \frac{1}{c(1+c)}\left[\hat{\lambda}_0 + \frac{\varepsilon}{c}\left(b_0 + \frac{b_1}{2\Omega}\right)\right], & \text{(two term)} \label{eq:lambda_final_expansion}
\end{align}
where $\lambda^{(0)}$ is a leading order approximation (keeping only the $O(\varepsilon^0)$ term) and $\lambda^{(1)}$ is a higher order approximation (keeping terms up to $O(\varepsilon^1)$).

Finally, to transform back to the nondimensional temperature $T$ anywhere in the domain, you take the product
\begin{align}
    T &= \lambda \phi.
\end{align}

\subsubsection*{Composite temperature field}
A composite temperature field, valid throughout the entire domain, can be constructed in the same way as that for the flow field, i.e. by forming composites between adjacent asymptotic regions. Splitting the domain (at the fin tips) into $y\geq 1$ and $y<1$, then we can construct a composite up to $O(\varepsilon)$ in each subdomain.

A gap--tip composite (restricted to $1\leq y \leq 1+c$) is given by
\begin{align}
    \phi_{\mathrm{gap-tip}} &= \underbrace{\phi_0(y) + \varepsilon \phi_1(y)}_{\text{gap region}} + \underbrace{\varepsilon \Phi_1 \left(X,\frac{y-1}{\varepsilon}\right)}_{\text{tip region}} - \underbrace{(1+c)\left(-(y - 1) - \frac{\varepsilon \log 2}{\pi} - \frac{\varepsilon}{2\Omega}\right)}_{\text{gap--tip overlap}}, \nonumber\\
    & \qquad 0<X<1, \label{eq:composite_temperature_upper}
\end{align}
bearing in mind that the gap solutions $\phi_0(y)$ and $\phi_1(y)$ are only known numerically, but $\Phi_1(X,Y)$ is (\ref{eq:Phi_1_solution}).

A tip--fin composite (restricted to $0\leq y <1$) between the tip and fin regions up to $O(\varepsilon)$ is given by
\begin{align}
    \phi_{\mathrm{tip-fin}} &=  \underbrace{\varepsilon \Phi_1 \left(X,\frac{y-1}{\varepsilon}\right)}_{\text{tip region}} + \underbrace{\varepsilon\tilde{\phi}_1(y)}_{\text{fin region}} + \underbrace{\frac{\varepsilon (1+c)}{2\Omega}}_{\text{overlap}}, \nonumber\\
    &= \varepsilon \Phi_1 \left(X,\frac{y-1}{\varepsilon}\right) - \frac{\varepsilon (1+c)}{2\Omega}(y-1),\qquad 0<X<1. \label{eq:composite_temperature_lower}
\end{align}

Also, in the fin itself, a second order composite (for $0\leq y \leq 1$) can be found and is given by (\ref{eq:second_order_fin_composite}).

\section{Nusselt numbers}
\label{sec:Nusselt_numbers}
In this section we present expressions for various Nusselt numbers as described in \cite{Sparrow1978}, and approximations for each in the narrow-fin-spacing limit $\varepsilon \to 0$.

First, the local Nusselt number on the fin is defined as
\begin{align}
    \mathrm{Nu}_\mathrm{f} &= \frac{q_\mathrm{f}^*H^*}{k(T_\mathrm{f}^* - T_\mathrm{b}^*)} = \frac{1}{1-\lambda \phi_{\mathrm{f}}} \frac{\lambda}{\varepsilon} \left.\frac{\partial \phi}{\partial X}\right|_{X=0} \quad \text{for }0<y<1. \label{eq:Nu_f_definition}
\end{align}
From the solution in the fin region, $\partial \phi /\partial X = O(\varepsilon^3)$ at most, hence $\mathrm{Nu}_f$ is only significant close to the fin tips. Substituting the solution there to $O(\varepsilon)$, we find
\begin{align}
    \mathrm{Nu}_\mathrm{f}(Y) &= \lambda_0 \left.\frac{\partial \Phi_1}{\partial X}\right|_{X=0} + O(\varepsilon) \\
    &= -\frac{\hat{\lambda}_0}{c\sqrt{\mathrm{e}^{-2\pi Y}-1}} + O(\varepsilon), \quad \text{for }Y<0. \label{eq:Nu_f_solution}
\end{align}
Notice that this leading order expression does not depend on the fin conductivity $\Omega$, and it is singular (with negative square root behaviour) and integrable at the fin tip, $Y\to 0$.

Second, the overall heat transfer is captured by the overall Nusselt number
\begin{align}
    \overline{\mathrm{Nu}} &= \frac{\bar{q}^*H^*}{k(T_{\mathrm{base}}^* - T_\mathrm{b}^*)}, \\
    \text{where }\bar{q}^* &= \frac{2\int_0^{H^*}q_\mathrm{f}^*\mathrm{d}y^* + \int_0^{S^*}q_\mathrm{base}^*\mathrm{d}x^*}{2H^* + S^*},
\end{align}
or in terms of nondimensional quantities,
\begin{align}
    \overline{\mathrm{Nu}} &= -\lambda \bar{q},\label{eq:Nu_bar_definition}\\
    \text{where }\bar{q} &=\frac{1}{2+\varepsilon} \left(2\int_0^1 -\left.\frac{\partial \phi}{\partial x}\right|_{x=0}\mathrm{d}y + \int_0^1 -\left.\frac{\partial \phi}{\partial y}\right|_{y=0}\mathrm{d}x\right). 
\end{align}
Here $\bar{q}$ is the nondimensional (with $\bar{q}^*$ the dimensional) average heat flux, in terms of $\phi$, out of the entire heat transfer surface available per period, i.e. two fins (height 1) and the base (width $\varepsilon$). This can be conveniently calculated from the problem statement for $\phi$. Integrating (\ref{eq:T_eq}) (with $T$ replaced by $\phi$) over the period domain $\mathcal{D}$ and applying the divergence theorem, 
\begin{align}
    \int_\mathcal{D}\lambda \mathcal{W}\phi~\mathrm{d}A &= 2\int_0^1 -\left.\frac{\partial \phi}{\partial x}\right|_{x=0}\mathrm{d}y + \int_0^1 -\left.\frac{\partial \phi}{\partial y}\right|_{y=0}\mathrm{d}x.
\end{align}
The left hand side reduces to simply $\varepsilon(1+c)$ by the definition (\ref{eq:lambda_phi_eq}) of the constant $\lambda$, and the right hand side equals $(2+\varepsilon)\bar{q}$. Hence $\bar{q}=\varepsilon(1+c)/(2+\varepsilon)$, and substituting into (\ref{eq:Nu_bar_definition}) gives
\begin{align}
    \overline{\mathrm{Nu}} &= -\lambda \frac{\varepsilon(1+c)}{2+\varepsilon}. \label{eq:Nu_bar_lambda_formula}
\end{align}
 
Substituting the approximation (\ref{eq:lambda_final_expansion}) for $\lambda$ in the limit $\varepsilon \ll 1$ leads to (at least) two levels of approximation for $\overline{\mathrm{Nu}}$, depending on the number of terms kept:
\begin{align}
    \overline{\mathrm{Nu}}^{(0)} &= -\frac{\varepsilon}{c}\frac{\hat{\lambda}_0}{(2+\varepsilon)} + O(\varepsilon^2), & \text{(leading order approx.)}  \label{eq:Nu_bar_leading_order}\\
    \overline{\mathrm{Nu}}^{(1)} &= -\frac{\varepsilon}{c} \frac{\displaystyle{\left(\hat{\lambda}_0 + \frac{\varepsilon}{c}\hat{\lambda}_1\right)}}{2 + \varepsilon} +O(\varepsilon^3),  & \text{(higher order approx.)} \label{eq:Nu_bar_higher_order}
\end{align}
Here $\overline{\mathrm{Nu}}^{(0)}$ is a leading order approximation (i.e. only the leading term of $\lambda$ is kept), and $\overline{\mathrm{Nu}}^{(1)}$ a higher order approximation (where both terms in $\lambda$ are kept). Note that we do not expand the denominator. We will show later that both of these approximations will have their advantages in practice, even though $\overline{\mathrm{Nu}}^{(1)}$ is strictly of a higher order.

Third, one can define a local Nusselt number on the base as
\begin{align}
    \mathrm{Nu}_{\mathrm{base}} &= \frac{q_\mathrm{base}^*H^*}{k(T_{\mathrm{base}}^* - T_\mathrm{b}^*)} = \lambda \left.\frac{\partial \phi}{\partial y}\right|_{y=0} \quad \text{for }0<X<1.
\end{align}
Substituting the solution (\ref{eq:phi_tilde_summary}) for $\phi$ in the fin region, and (\ref{eq:lambda_final_expansion}) for $\lambda$, and retaining different orders in $\varepsilon$ leads to the approximations:
\begin{align}
    \mathrm{Nu}_{\mathrm{base}}^{(0)} &=  -\frac{\varepsilon\hat{\lambda}_0}{2c\Omega} + O(\varepsilon^2), & \text{(leading order approx.)} \label{eq:Nu_base_leading_order}\\
    \mathrm{Nu}_{\mathrm{base}}^{(1)} &= -\frac{\varepsilon}{2c\Omega}\left(\hat{\lambda}_0 + \frac{\varepsilon}{c}\hat{\lambda}_1\right)\left(1 - \frac{\varepsilon}{2\Omega}\right) + O(\varepsilon^3), & \text{(higher order approx.)}    \label{eq:Nu_base_higher_order}
\end{align}
This is constant in $X$ to the orders given, and hence the base-averaged Nusselt $\overline{\mathrm{Nu}}_{\mathrm{base}}$ takes the same value.

\section{Comparison to numerical solution}
\label{sec:results_comparison}
To assess the validity of the solutions derived throughout the paper, we compare to numerical solutions of the full hydrodynamic and thermal problems for arbitrary values of the parameters (in particular, arbitrary fin spacing-to-height ratio $\varepsilon$). For a detailed discussion of the impact of $\varepsilon$ on the general heat transfer behaviour, we refer to \cite{Sparrow1978}. Here we focus on the new asymptotic results, and what we can learn from them.

\subsection{Numerical methodology}
To solve the full nonlinear two-dimensional problem we employ a domain decomposition and pseudospectral (Chebyshev) collocation methods suited to boundary value problems. The approach is similar to that used by \cite{Game_et_al18,Mayer2021} in a different flow context (superhydrophobic surfaces), and so we give a brief description here with further details given in Supplementary Material.

One-half period is split into 3 subdomains: 2 fluid (two-dimensional) domains and one fin (one-dimensional) domain, and each are discretized into $N+1$ points in $x$ (excluding the fin domain) and $M+1$ points in $y$ using the Gauss--Lobatto spacing for spectral accuracy \cite{TrefethenSpectralMethods}. This clusters grid points close to domain boundaries (hence even more so for domain corners), and since the fluid domain is subdivided at $y=1$, many points are clustered close to the singular point ($(x,y)=(0,1)$) as it sits at the corner of two domains. This helps ensure numerical resolution of the large gradients close to this point where the boundary conditions change type. Continuity of velocity, temperature, shear stress and heat flux were imposed at common boundaries between the fluid subdomains.

The velocity problem discretizes into a linear system that is inverted in MATLAB. This is fed into the thermal problem, employing the same mesh (with the addition of the 1D fin domain), which is nonlinear and solved in an iterative fashion. The iteration method is similar to that of \cite{Sparrow1978,Karamanis-16}, whereby the advection (left hand side) in (\ref{eq:T_eq}) is assumed known from the previous iteration. In this way, the problem to update $\phi$ is linear and decoupled from $\lambda$, which is subsequently updated using (\ref{eq:lambda_phi_eq}). Convergence is rapid, with less than 10 iterations required to achieve a relative change in $\lambda$ of less than 0.01\%.

Convergence of the spatial discretization was confirmed by doubling $N$ and $M$ until $\lambda$ changed by less than 0.5\%. For all of the results shown, typically $N=20$ was sufficient, with $M$ chosen as follows. Since we are concerned with small spacing-to-height ratios $\varepsilon$, we found it important to increase $M$ (vertical grid points per subdomain) as $\varepsilon$ decreases. Indeed, we related $M$ to $N$ via the scaling $M = N/(2\varepsilon)$, until a maximum value of $M=100$. This lead to good results down to an $\varepsilon$ value of $0.025$. To decrease $\varepsilon$ further without unreasonable computational effort, one could introduce additional subdomains (e.g. close to the fin tip) to cluster the grid points more efficiently, or instead subtract the local form of the singularity at $(x,y)=(0,1)$ (as in \cite{Game_et_al18}). However, this was not the focus of our study.

The numerical results for $\overline{\mathrm{Nu}}$ were compared (see Fig. \ref{fig:Nu_bar}) to those of \cite{Sparrow1978}, with differences on the order of one percent or less---the discrepancies likely due to the coarser mesh of Sparrow et al. (who took at most 50 equally spaced points in the $y$ direction).

\subsection{Velocity and temperature fields: $w(x,y),\phi(x,y)$}

\begin{figure}
    \centering
    \includegraphics[width=0.90\textwidth, trim = {1.5cm 0 1.5cm 0}, clip]{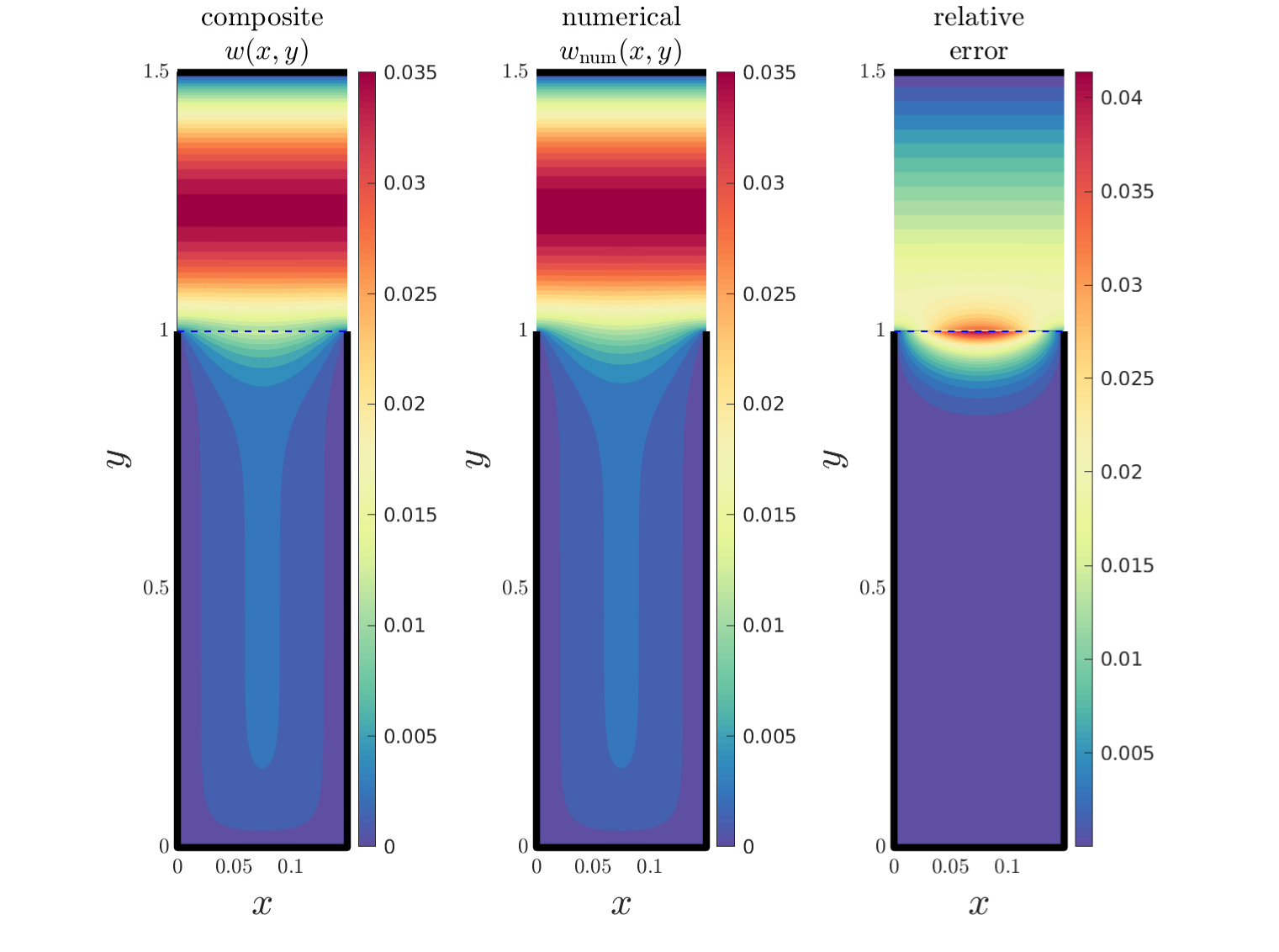}
    \caption{The asymptotic piece-wise composite solution (\ref{eq:composite_velocity_upper})-(\ref{eq:composite_velocity_lower}) for the velocity field $w(x,y)$ in one period (left panel), compared to the numerical velocity $w_\mathrm{num}$ (middle panel), with the relative error $|w(x,y) -w_\mathrm{num}(x,y)|/\mathrm{max}|w_\mathrm{num}|$ (right). Geometric parameters are $\varepsilon=0.15$ and $c=0.5$. The asymptotic solution consists of two composites: one valid for $y\geq 1$, and one for $y<1$, and the separating line ($y=1$) is shown as a dashed blue line. The fins (at $0\leq y \leq 1$, $x=0,1$) and base are shown in black.}
    \label{fig:composite_velocity}
\end{figure}
\begin{figure}
    \centering
    \includegraphics[width=0.9\textwidth, trim = {1.5cm 0 1.5cm 0}, clip]{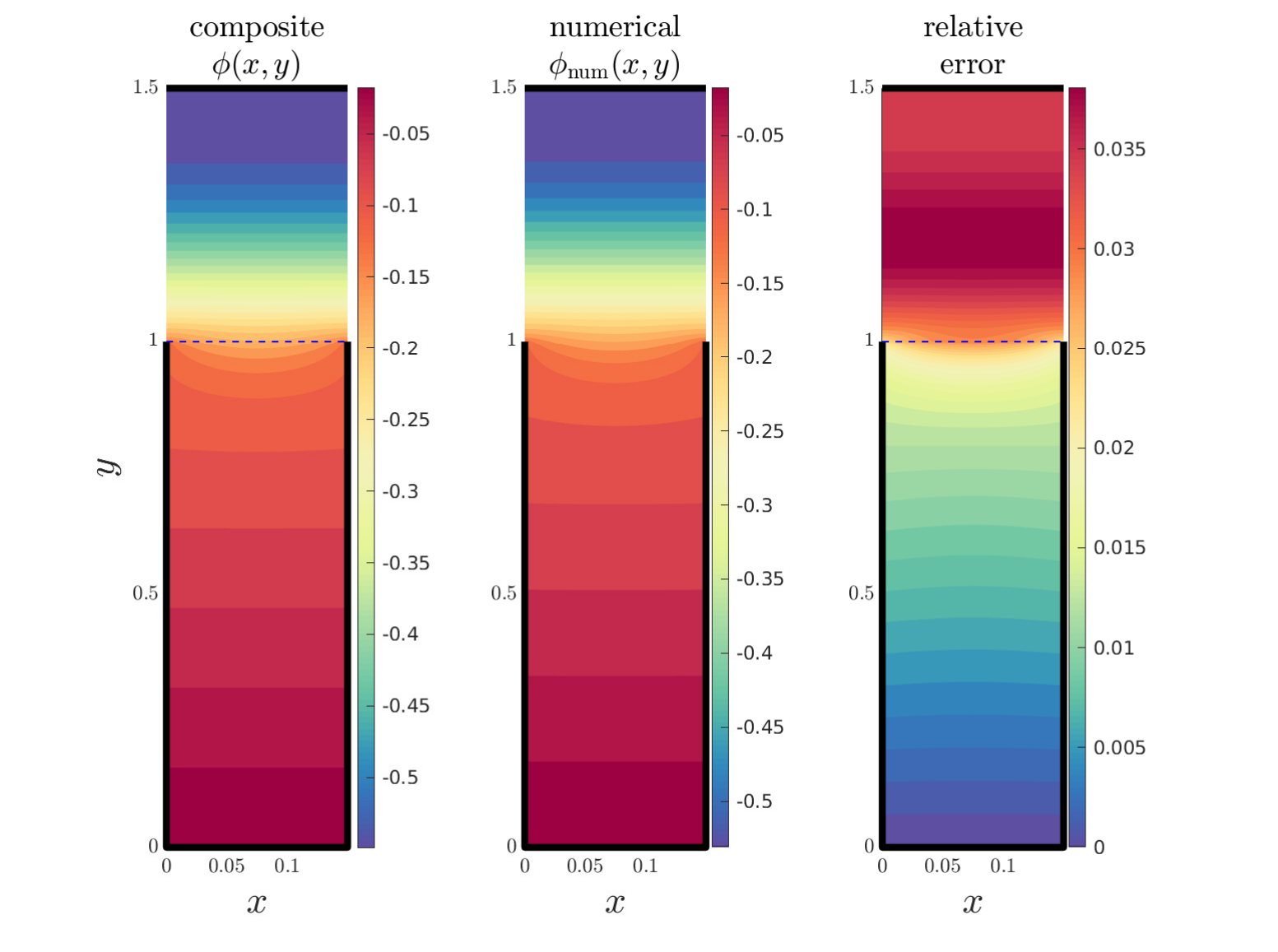}
    \caption{The asymptotic piece-wise composite solution (\ref{eq:composite_temperature_upper})-(\ref{eq:composite_temperature_lower}) for the scaled temperature field $\phi(x,y)=T/\lambda$ in one period (left panel), compared to the numerical temperature (middle), with the relative error $|\phi(x,y) -\phi_\mathrm{num}(x,y)|/\mathrm{max}|\phi_\mathrm{num}|$ (right panel). Here $\varepsilon=0.15$, $c=0.5$, and $\Omega =1$. See caption for Fig. \ref{fig:composite_velocity}.}
    \label{fig:composite_temperature}
\end{figure}

We begin by comparing the asymptotic composite solutions for $w(x,y)$ ((\ref{eq:composite_velocity_upper})-(\ref{eq:composite_velocity_lower})) and $\phi(x,y)$ ((\ref{eq:composite_temperature_upper})-(\ref{eq:composite_temperature_lower})) to the numerical solutions. Fig. \ref{fig:composite_velocity} compares $w(x,y)$ for the case $\varepsilon=0.15$ (on the upper end of the practical range) and $c=0.5$. The structure of the flow is captured excellently by the asymptotic solution: the bulk of the flow occurs above the fins in an almost parabolic profile, and only weakly penetrates down into the space between the fins (where the velocity is an order of magnitude lower). The composite is actually discontinuous on the line $y=1$, due to there being $O(\varepsilon^2)$ terms in $y<1$ and not in $y\geq 1$. Consequently, this is where the relative error is the greatest (see right of Fig. \ref{fig:composite_velocity}), albeit still $\lesssim 4\%$.

Fig.~\ref{fig:composite_temperature} compares $\phi(x,y)$ for the same geometry, and for a fin conductivity parameter $\Omega = 1$. Again, the field structure is captured, where the heat transfer is predominantly 1D conduction from the base until close to the fin tips where the conduction becomes 2D (from the fin to the fluid), followed by a 1D convection behaviour (and nonlinear temperature profile) above the fins. The relatively low value of the fin conductivity parameter, $\Omega=1$, was chosen to exhibit the 1D conduction clearly, but in practice $\Omega$ will generally be larger \cite{Karamanis-19b}. The relative error is also small ($\lesssim 3.5\%$), and is greatest at the top of the domain due to errors in each region accumulating as you move away from the isothermal condition $\phi=0$ on $y=0$. Unlike the velocity composite, the temperature composite is indeed continuous at $y=1$.

We have shown here the scaled temperature $\phi$, not the ``nondimensional temperature" $T$, which is related via $T = \lambda \phi$. This was done in an attempt to compare the temperature \emph{distributions}. The overall magnitude is affected greatly by the approximation to $\lambda$, which we assess separately.

\subsection{The constant $\lambda$}
\begin{figure}
    \centering
    \includegraphics[width=1\textwidth]{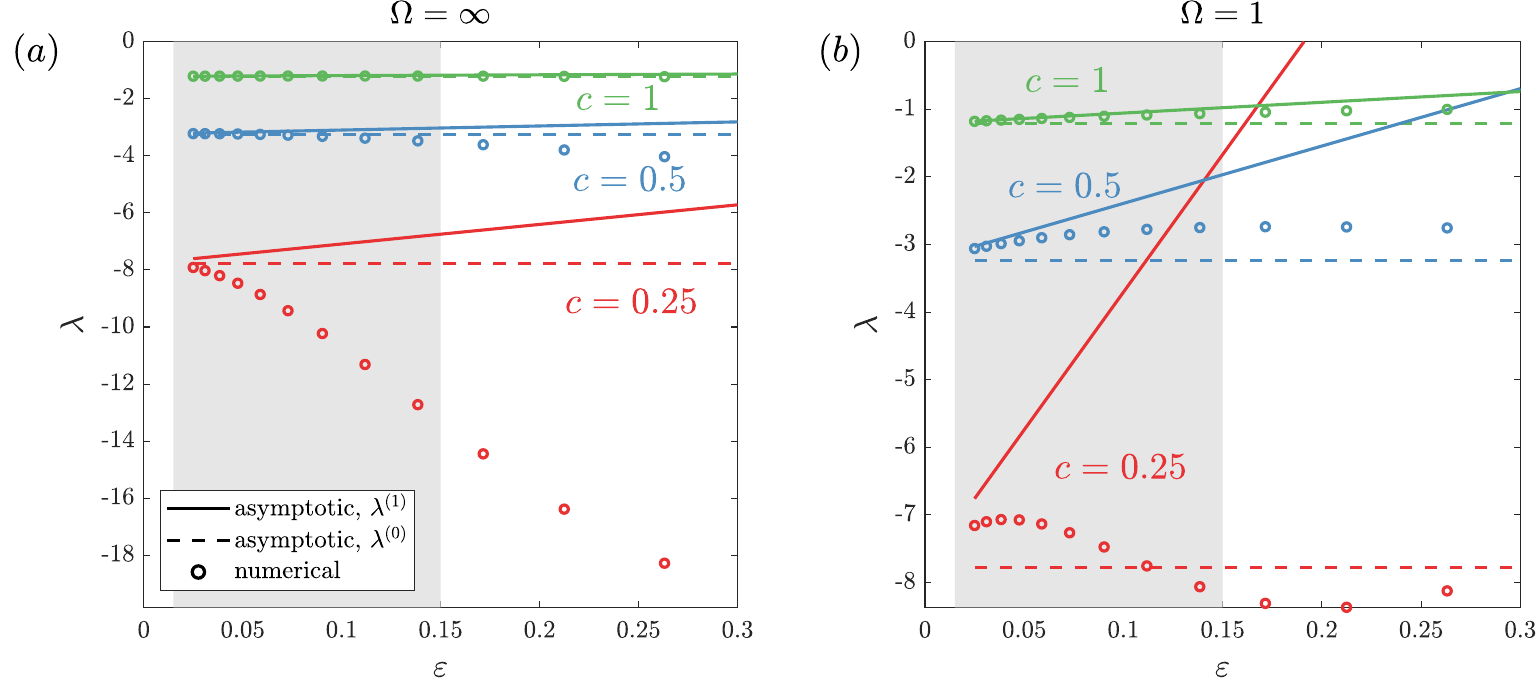}
    \caption{The solution constant $\lambda$, comparing the asymptotic solutions  for $\varepsilon \ll1 $ to the numerical solution. Dashed lines are using only the leading order (\ref{eq:lambda_leading_order}) and solid lines are the full (two term) approximation (\ref{eq:lambda_final_expansion}). Shaded region shows the realistic range $0.015 < \varepsilon < 0.15$ of fin spacings applicable to manufacturable heat sinks \cite{Iyengar-07}.}
    \label{fig:lambda_comparison}
\end{figure}
\begin{figure}
    \centering
    \includegraphics[width=0.95\textwidth, trim = {1.6cm 0 2.5cm 0}, clip]{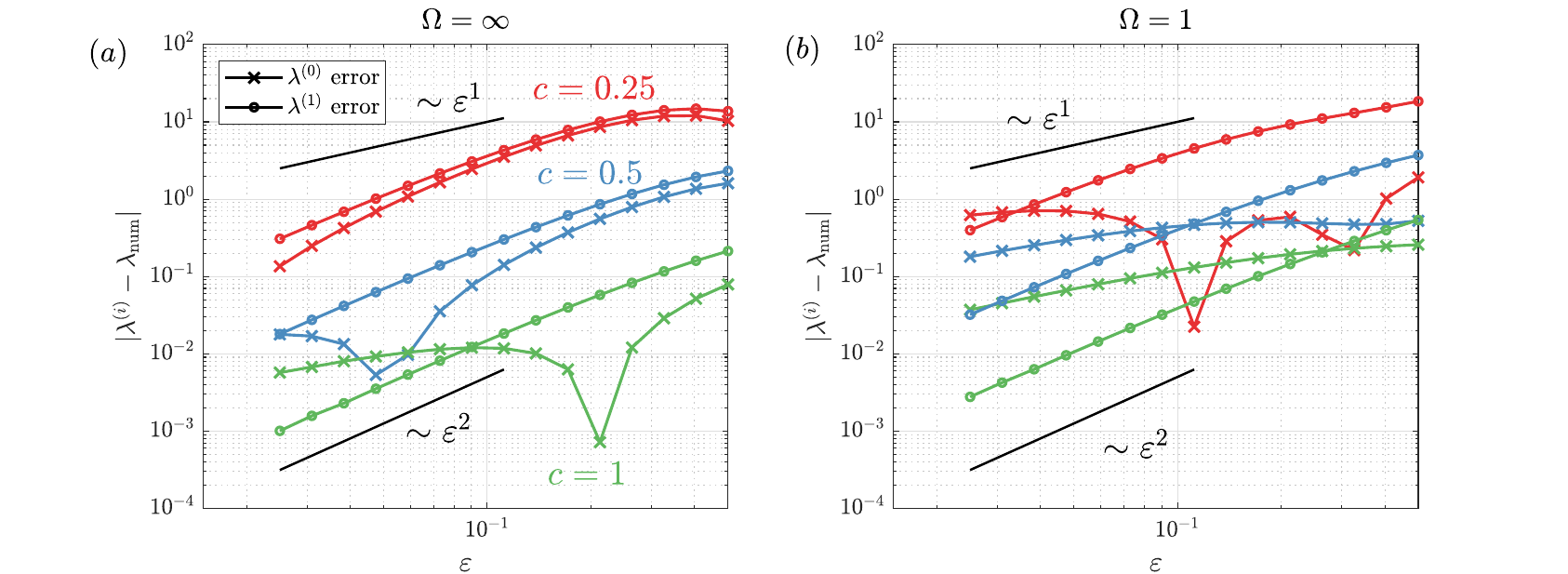}%
    \caption{Absolute error of the asymptotic approximations for the constant $\lambda$, compared to the numerical solution. The error is shown for two levels of approximation, $\lambda^{(0)} = \lambda_0$ and $\lambda^{(1)} = \lambda_0 + \varepsilon\lambda_1$. Values of $c$ and $\Omega$ are indicated.}
    \label{fig:lambda_error}
\end{figure}
In this section we compare the asymptotic approximation(s) for the solution constant $\lambda$ with the numerical solution. Recall that $\lambda$ can be interpreted as the exponential decay rate in the streamwise direction ($z$) of the temperature profile towards the isothermal base temperature. It also directly gives the overall Nusselt number (\ref{eq:Nu_bar_lambda_formula}).
Figure \ref{fig:lambda_comparison} compares two approximations for $\lambda$ against numerical solutions for a range of $\varepsilon$ (down to $\varepsilon=0.025$), and selected $c$ and $\Omega$ values. In all cases, the approximation $\lambda^{(0)}$ captures the limiting value of $\lambda_\mathrm{num}$ as $\varepsilon\to 0$ (as it should), and  $\lambda^{(1)}$ captures the value and slope (as it should). However, even though the slope is captured, as $c$ decreases it is clear that the range of $\varepsilon$ for which $\lambda^{(1)}$ is a good approximation also decreases. This is expected, since the asymptotic approximation requires $\varepsilon \ll1$ but also $\varepsilon \ll c$, i.e. that the fin spacing is small compared to the clearance. Hence the assumption breaks down if, e.g. $c = O(\varepsilon)$ (as seen for $f\mathrm{Re}$ in \cite{Miyoshi2024}), clear from (\ref{eq:lambda_final_expansion}) since the correction is proportional to $\varepsilon / c$. Additionally, the accuracy of $\lambda^{(1)}$ decreases as $\Omega$ decreases, also seen from the correction. 

Indeed, from the expansion (\ref{eq:lambda_final_expansion}) one can summarise the validity requirements as $\varepsilon/c \ll 1$ and $\varepsilon / (c \Omega)\ll 1$, since otherwise the correction is comparable to the leading term. The first condition, $\varepsilon/c \ll 1$, is purely a geometric requirement necessary for there to be a gap region above the fins where the flow and temperature depends only on $y$. The second condition, $\varepsilon / (c \Omega)\ll 1$, is a further constraint that applies to the conductivity of the fin. The leading order temperature in the fin is $T_\mathrm{f} = -\hat{\lambda}_0 \varepsilon y / (c\Omega) + \cdots $, implying that the dimensionless temperature drop $\Delta T_\mathrm{f}$ across the height of the fin is $\Delta T_\mathrm{f} \sim \varepsilon / (c\Omega)$. Hence the constraint above implies $\Delta T_\mathrm{f} \ll 1$, or in dimensional terms,
\begin{align}
    T_\mathrm{base}^* - T_\mathrm{f, tip}^* \ll T_\mathrm{base}^* - T_\mathrm{b}^*,
\end{align}
i.e., the temperature drop from the base to the fin tip is small compared to the drop from the base to the fluid bulk. If the fin is not conductive enough (due to solid/fluid conductivity or fin thickness) then the temperature at the tip of the fin becomes comparable to the bulk temperature $T_\mathrm{b}^*$. Then the heat transfer between the fin and the bulk cannot predominantly occur near the tip: a significant portion must also occur further down the fin---and thus advection there (not just in the gap region above) would need to be considered.    

Moving on, a surprising result is the accuracy of the leading order solution $\lambda^{(0)}$. When the fins are highly conductive (say $\Omega=\infty$, Fig. \ref{fig:lambda_comparison}$(a)$), $\lambda^{(0)}$ and $\lambda^{(1)}$ are similar approximations, but $\lambda^{(0)}$ is slightly closer (for larger $\varepsilon$) to the numerics as $c$ is decreased. This is further exaggerated as the fins becomes less conductive (Fig. \ref{fig:lambda_comparison}$(b)$). Mathematically, it is clear that this is due to the strong nonmonotonic dependence of $\lambda$ on $\varepsilon$: the better accuracy of $\lambda^{(1)}$ for small $\varepsilon$ results in worse accuracy for larger $\varepsilon$. However, this significant advantage of $\lambda^{(0)}$ over $\lambda^{(1)}$ seems to occur mainly outside of the realistic range of fin spacings ($0.015 < \varepsilon < 0.15$), shown in grey.

The above points are made more apparent by looking at the absolute error of $\lambda^{(0)}$ and $\lambda^{(1)}$ in Fig. \ref{fig:lambda_error}. The errors of $\lambda^{(0)}$ and $\lambda^{(1)}$ show slopes of 1 (indicating $O(\varepsilon)$) and 2 (indicating $O(\varepsilon^2)$), respectively, as $\varepsilon \to 0$. For moderately large $\varepsilon$, $\lambda^{(0)}$ typically shows the smaller error, but below some critical value of $\varepsilon$ (which increases with $c$), $\lambda^{(1)}$ always becomes more accurate. This has consequences for the $\overline{\mathrm{Nu}}$ approximations, discussed in section \ref{sec:results_Nu_bar}.

\subsection{Local Nusselt number on the fin surface}
\begin{figure}
    \centering
    \includegraphics[width=0.9\textwidth, trim = {0.1cm 0 2cm 0}, clip]{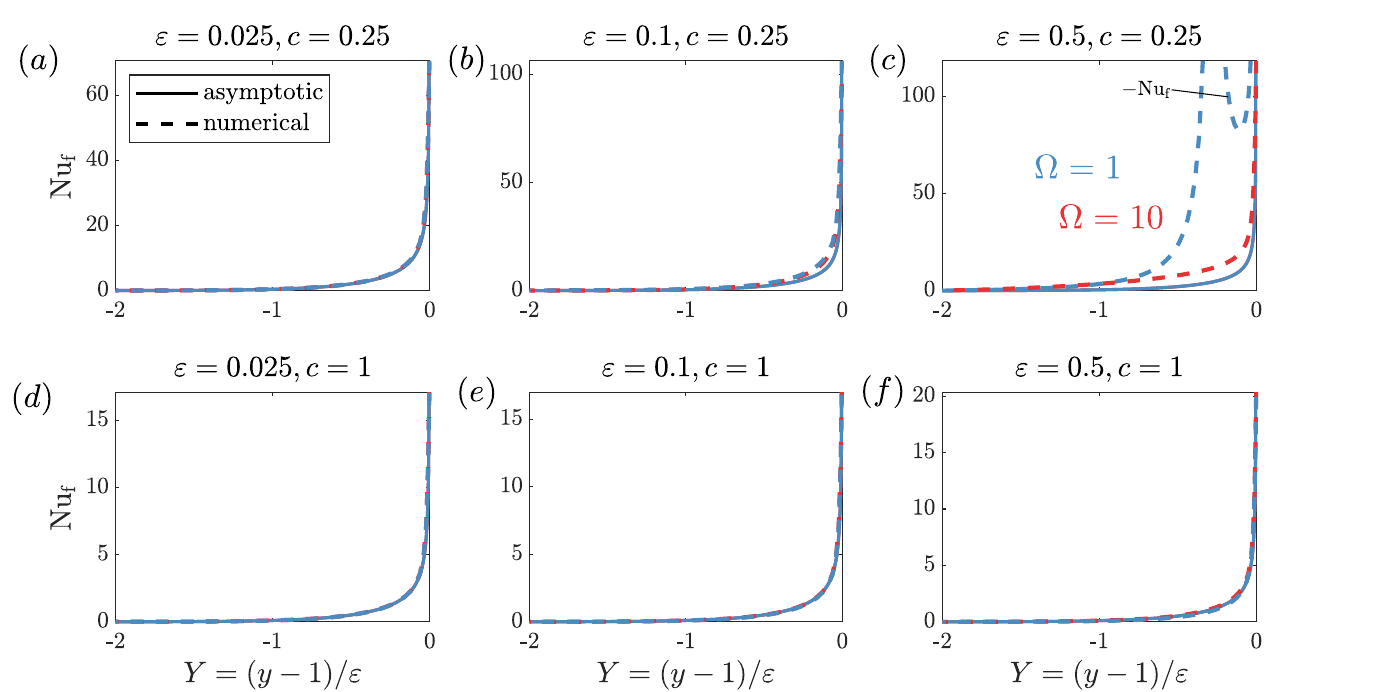}
    \caption{The local Nusselt number on the fin surface, $\mathrm{Nu}_\mathrm{f}(Y)$ as a function of the tip variable $Y = (y-1)/\varepsilon$, comparing asymptotic (given by (\ref{eq:Nu_f_solution})) and numerical solutions. The values of $c,\varepsilon$ and $\Omega$ are indicated. Different solutions are only distinguishable in $(b)$, $(c)$.}
    \label{fig:Nu_f}
\end{figure}

In this section we assess the approximation of local quantities along the surface of the fin. We begin with the local Nusselt number $\mathrm{Nu}_\mathrm{f}$, given numerically by definition (\ref{eq:Nu_f_definition}), and approximated by (\ref{eq:Nu_f_solution}). Fig. \ref{fig:Nu_f} compares $\mathrm{Nu}_\mathrm{f}$ for a range of $\varepsilon$, $c$, and $\Omega$ values. It is plotted in terms of the tip variable $Y$, and the behaviour is reminiscent of that in \cite{Sparrow1978}. Our formula shows that it is singular ($\mathrm{Nu}_\mathrm{f}=O(|Y|^{-1/2})$) as $Y \to 0$, and decays very quickly ($\mathrm{Nu}_\mathrm{f}=O(\mathrm{e}^{-\pi |Y|})$) as you move away from the tip, $Y\to -\infty$. The singularity is expected from the change of thermal boundary condition there from a Dirichlet one ($T=T_\mathrm{f}, y<1$) to a Neumann one ($\partial T/\partial x=0, y>1$), and the asymptotic solution captures the behaviour excellently. Indeed, the shape barely changes as the parameters are varied (its magnitude scales with $c$), only when the asymptotic assumption breaks down, i.e. the case $c=0.25 < \varepsilon =0.5$. As remarked by Sparrow et al., this behaviour is very far from the common assumption of a ``constant heat transfer coefficient" along the fin surface, and it is consistent across parameters, even when the fin conductivity is low ($\Omega = 1$).

\begin{figure}
    \centering
    \includegraphics[width=0.9\textwidth, trim = {0.1cm 0 2cm 0}, clip]{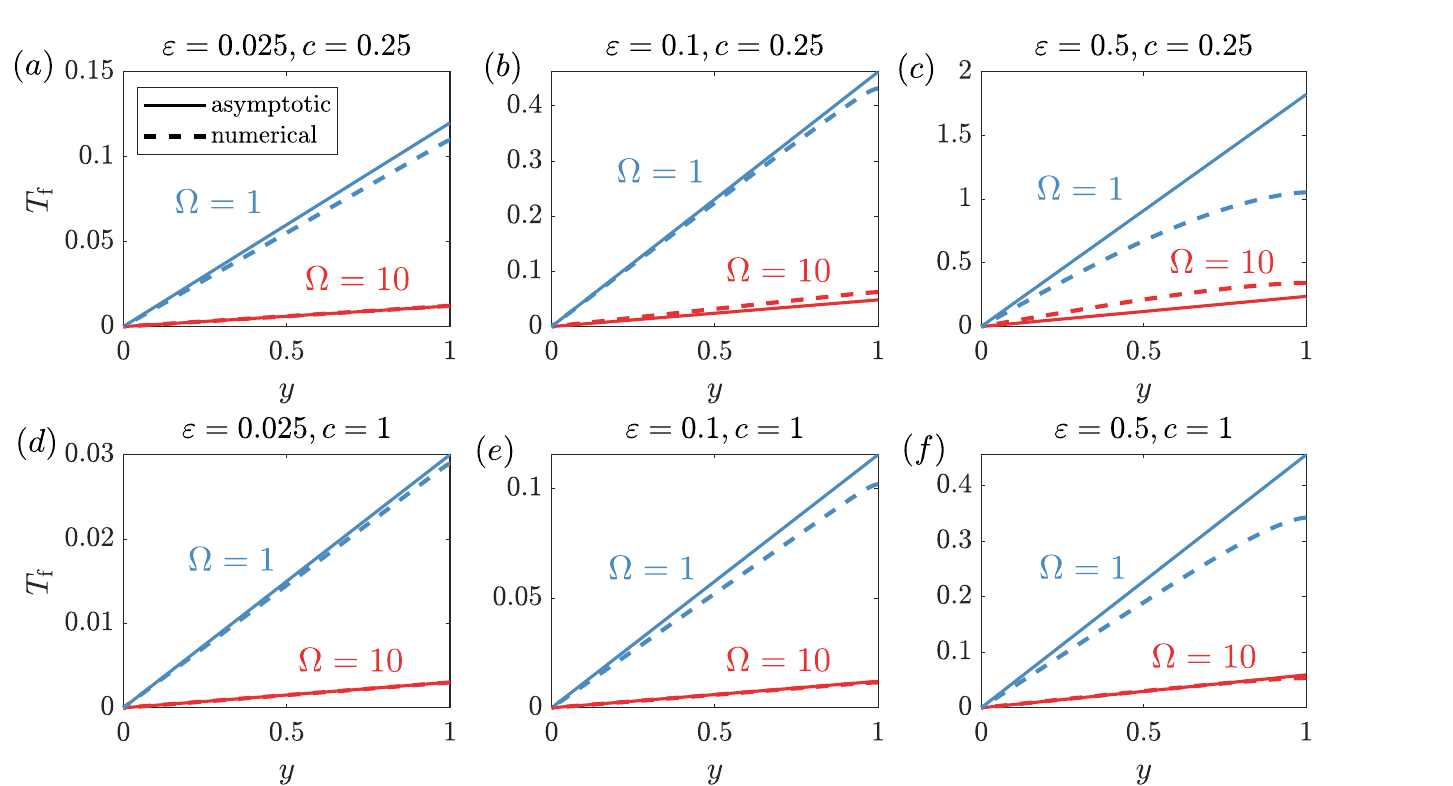}
    \caption{The temperature along the fin $T_\mathrm{f}(y)$ comparing the numerical solution (spectral collocation), and the asymptotic solution $T_\mathrm{f} \sim \lambda_0 \tilde{\phi}_{\mathrm{f}}$ in the fin region ($0 \leq y$ and $1-y \gg \varepsilon$), where $\tilde{\phi}_\mathrm{f}$ is (\ref{eq:phi_tilde_summary}). The values of $c,\varepsilon$ and $\Omega$ are indicated.}
    \label{fig:T_f}
\end{figure}

The nondimensional temperature $T_\mathrm{f} = (T_\mathrm{f}^*-T_{\mathrm{base}}^*)(T_{\mathrm{b}}^* - T_{\mathrm{base}}^*)$, for the same parameter cases, is shown in Fig. \ref{fig:T_f}. The linear 1D conduction behaviour in $y$, exhibited by the asymptotic solution (\ref{eq:phi_tilde_summary}), is observed in most cases. It departs from linear close to the tip ($y = 1$), in a small region that widens with $\varepsilon$. The agreement worsens as $\varepsilon$ increases, and $c$ or $\Omega$ decreases. Note that we used the leading order approximation for $\lambda = \lambda_0 + \ldots$ here, since moderate accuracy for a wider range of $\varepsilon$ was necessary for illustration.


\subsection{Overall Nusselt number $\overline{\mathrm{Nu}}$}
\label{sec:results_Nu_bar}
\begin{figure}
    \centering
    \includegraphics[width=1\textwidth]{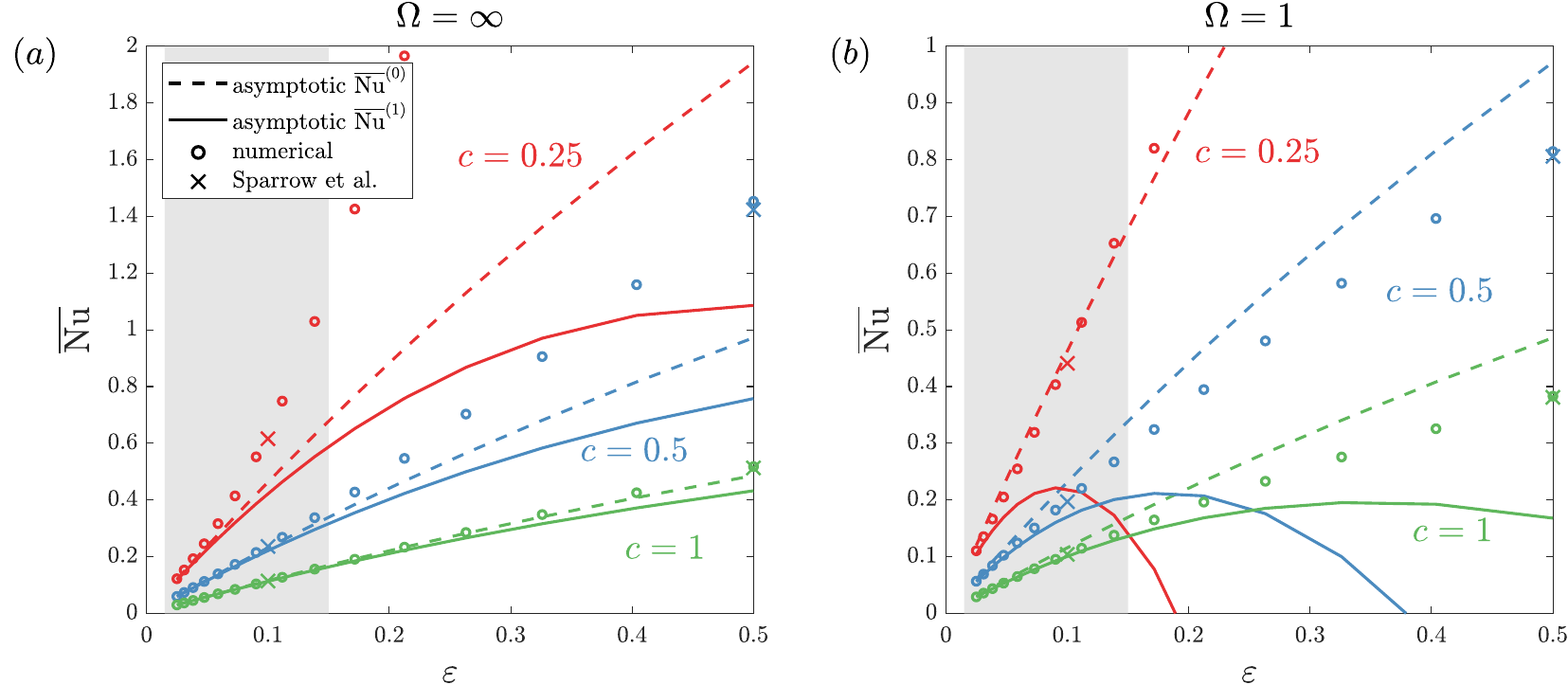}
    \caption{The overall average Nusselt number $\overline{\mathrm{Nu}}$, comparing the asymptotic solutions (\ref{eq:Nu_bar_leading_order})-(\ref{eq:Nu_bar_higher_order}) for $\varepsilon \ll1 $ to the numerical solution. Values from \cite{Sparrow1978} are also shown, where available. Shaded region shows the realistic range $0.015 < \varepsilon < 0.15$ applicable to manufacturable heat sinks.}
    \label{fig:Nu_bar}
\end{figure}

Finally, we compare the approximations for the overall Nusselt number, which gives a measure of the overall efficacy of the heat transfer to the fluid. Fig. \ref{fig:Nu_bar} compares (\ref{eq:Nu_bar_leading_order})-(\ref{eq:Nu_bar_higher_order}) to the numerical solutions for the same parameters as in Fig. \ref{fig:lambda_comparison}. Since $\overline{\mathrm{Nu}}$ follows directly from $\lambda$, the approximations exhibit similar behaviour and regions of validity as those for $\lambda$, but most notably $\overline{\mathrm{Nu}}$ contains an additional factor of $\varepsilon$ and thus $\overline{\mathrm{Nu}} \to 0$ as $\varepsilon \to 0$. Similar to $\lambda$, both approximations show good agreement with the numerics, but $\overline{\mathrm{Nu}}^{(1)}$ shows excellent agreement for smaller values of $\varepsilon$, and $\overline{\mathrm{Nu}}^{(0)}$ shows moderate agreement over a larger range of $\varepsilon$. Notably, $\overline{\mathrm{Nu}}^{(0)}$ remains positive whereas $\overline{\mathrm{Nu}}^{(1)}$ can erroneously become negative far outside its range of validity.

The values of $\overline{\mathrm{Nu}}$ here in the realistic $\varepsilon$ range are remarkably low compared to the typical values without tip clearance ($c=0$), where $\overline{\mathrm{Nu}}\approx 2 - 34$ \cite{Sparrow1978}. Thus, to avoid gross overestimation, it is crucial to not use such values when there is tip clearance, but the formulas provided here instead. 

\begin{figure}
    \centering
    \includegraphics[width=1\textwidth]{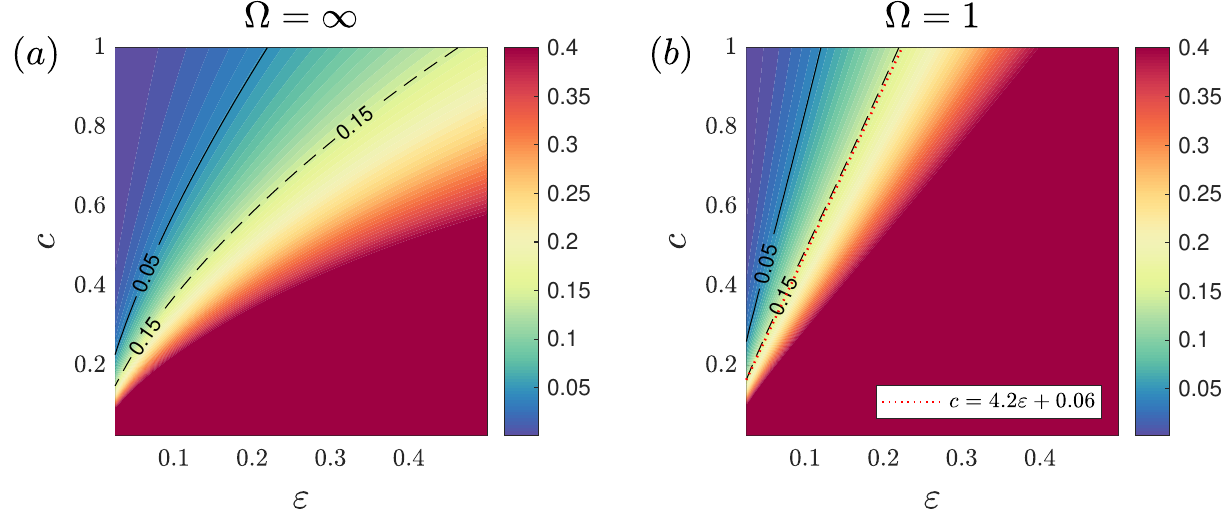}
    \caption{Relative error of approximation $\overline{\mathrm{Nu}}^{(1)}$ (see (\ref{eq:Nu_bar_higher_order})), compared to the numerical solution, i.e. contours of $|\overline{\mathrm{Nu}}^{(1)}-\overline{\mathrm{Nu}}_\mathrm{num}|/\overline{\mathrm{Nu}}_\mathrm{num}$ for $\varepsilon\in [0.025,0.5]$, $c\in [0.025, 1]$. The marginal boundaries where the error is 0.05 and 0.15 (5\% and 15\%) are shown as black lines and labelled.}
    \label{fig:error_contours_Nu_bar_1}
\end{figure}

The accuracy of $\overline{\mathrm{Nu}}^{(1)}$ is perhaps better shown in Fig. \ref{fig:error_contours_Nu_bar_1} where the relative error is plotted in the $(c,\varepsilon)$-plane for $\varepsilon\in [0.025,0.5]$, $c\in [0.025, 1]$. Error is lowest when $\varepsilon$ is small and $c$ and $\Omega$ are large, consistent with the asymptotic requirements that $\varepsilon / c \ll 1$ and $\varepsilon / (c \Omega) \ll 1$. Also shown are black marginal lines, above which the relative error is less than 5\% or 15\%. For the range plotted, the error of $\overline{\mathrm{Nu}}^{(1)}$ was found to be $<15\%$ in the following approximate region (provided $\Omega \geq 1$): 
\begin{align}
    c \geq 4.2 \varepsilon + 0.06. \label{eq:c_range}
\end{align}

\begin{figure}
    \centering
    \includegraphics[width=1\textwidth]{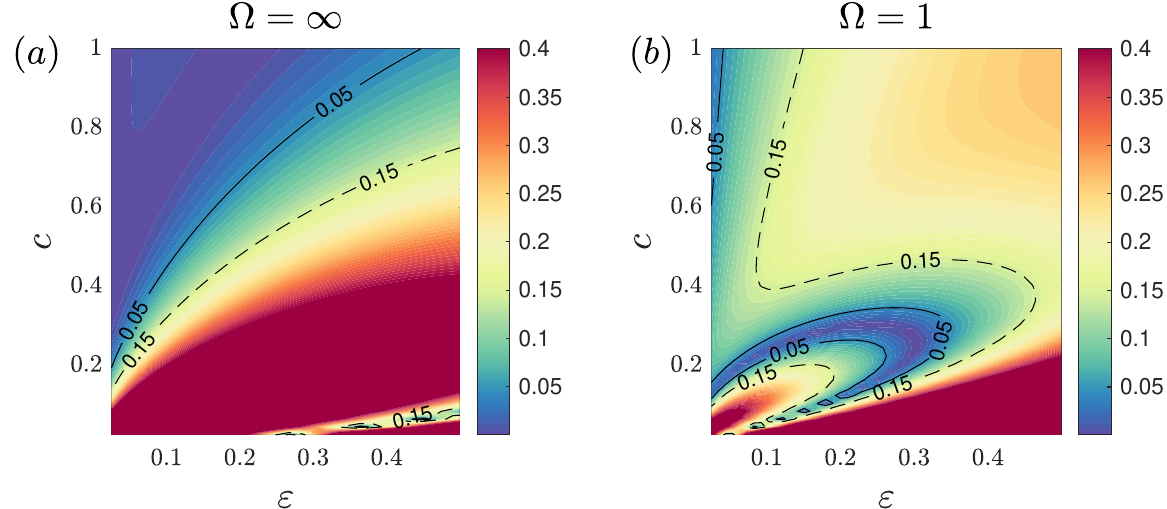}
    \caption{Relative error of approximation $\overline{\mathrm{Nu}}^{(0)}$ (\ref{eq:Nu_bar_leading_order}). See caption for for Fig. \ref{fig:error_contours_Nu_bar_1}.}
    \label{fig:error_contours_Nu_bar_0}
\end{figure}

For completeness, we also plot the error of $\overline{\mathrm{Nu}}^{(0)}$ in Fig. \ref{fig:error_contours_Nu_bar_0}. The behaviour is far less predictable, but the regions where the error is $<15\%$ are generally larger than for $\overline{\mathrm{Nu}}^{(1)}$. In particular, when $\Omega = 1$ and $0.2 < c< 0.4$, this region extends up to $\varepsilon < 0.45$. Hence $\overline{\mathrm{Nu}}^{(0)}$ can have particular use cases over $\overline{\mathrm{Nu}}^{(1)}$, but these cases may be difficult to predict and require trial and error to determine.



\section{Conclusions}
\label{sec:conclusions}
In this paper, we considered the convective heat transfer problem from a longitudinal-finned heat sink (isothermal base) with tip clearance---a classic and ubiquitous problem in the thermal management of electronics. We considered an analytical approach to the mathematical problem formulated and solved numerically by \cite{Sparrow1978}. In particular, we presented a detailed asymptotic solution for the flow and (coupled, i.e. conjugate) temperature fields in the fluid and the fins, in the limit of small fin spacing, $\varepsilon \ll 1$. 

We used the method of matched asymptotic expansions, and decomposed the flow domain into regions where different transport processes are important: (i) a gap region (above the fins) where the flow is the fastest but 1D, driving a 1D convection profile; (ii) a tip region (close to the fin tips) where the flow and temperature fields are 2D (purely diffusive); (iii) a fin region (down between the fins) where the flow is almost stagnant, and so the thermal problem is purely 1D conduction and governed by relative conductivity of the fin. The structure elaborates on the insights of Sparrow \emph{et al.} on why the heat transfer suffers when there is clearance. Because the heat transfer predominantly occurs close to the fin tips, the use of uniform heat transfer coefficients is wholly inappropriate: instead the local coefficient is singular at the tip and decays to a negligible value after approximately 1 fin spacing from the tip.

The 1D convection problems in the gap region at each order only need to be solved once, and this is trivially done numerically. Remarkably, explicit solutions are found for the local heat transfer coefficient (or Nusselt number $\mathrm{Nu}_\mathrm{f}$); the fin and fluid temperature throughout the entire fin and tip regions; the heat flux along the length of the fin; and the overall Nusselt number ($\overline{\mathrm{Nu}}$). Therefore, one does not have to rely purely on numerical solutions, or crude (and inaccurate) uniform assumptions, for this relevant heat transfer problem anymore. These solutions are compared to numerical solutions of the full coupled problem for arbitrary $\varepsilon$ and are found to agree well for small $\varepsilon$ in the realistic range ($0.015 \leq \varepsilon \leq 0.15$). In general, they are valid under the parametric assumptions $\varepsilon \ll1$, $\varepsilon / c \ll1$ and $\varepsilon / (c\Omega) \ll 1$. The first two are geometric assumptions, and the third ensures that the temperature drop along the fin is small compared to the overall driving temperature difference between the base ($T^*_\mathrm{base}$) and the fluid bulk ($T^*_\mathrm{b}$). Thus, accuracy decreases if $\varepsilon$ increases, $c$ decreases, or $\Omega$ decreases.

The leading order formula for the overall heat transfer ($\overline{\mathrm{Nu}}$) shows good accuracy for a moderate range of $\varepsilon$. The higher order formula gives better accuracy when $\varepsilon$ is small, but at the expense of worse accuracy for moderate $\varepsilon$. Nonetheless, the error of the higher order formula was found to be $<15\%$ for all $\varepsilon$ we considered, provided $c\geq 4.2\varepsilon + 0.06$ and $\Omega \geq 1$.

We focused here on the scenario with tip clearance ($c>0$). The case where the heat sink is fully shrouded, without tip clearance ($c=0$) is another very relevant one. The transport structure is significantly different since the location of highest flow velocity occurs between the fins, and hence advection there appears at leading order and cannot be neglected. The solution is thus not simply the limit $c\to 0$ of the solution presented here. The $c=0$ case will be published in future work.

In summary, our companion paper by \cite{Miyoshi2024} provided analytical results for the Poiseuille number through rectangular ducts with two (facing) surfaces having mixed (no-slip and shear-free) boundary conditions, which previously had not been done. Extending this configuration to a diabatic problem, the present study provides analytical results for conjugate Nusselt numbers through such ducts having (facing) surfaces where the conjugate heat transfer problem has been resolved.

\appendix
\section{Leading order heat flux on the fin surface}
\label{sec:heat_flux_derivation}

Calculation of the normal derivative $\partial W_1/\partial X$ on the fin surface $X=0$, $Y<0$, is of interest to the thermal problem in the tip region. In particular, we are interested in the local heat flux out of the fin,
\begin{align}
    -\left.\frac{\partial \Phi_1}{\partial X}\right|_{X=0} &= \frac{2(1+c)}{c} \left.\frac{\partial W_1}{\partial X}\right|_{X=0}.
\end{align}
The solution for $W_1$ is determined as the imaginary part of a function $h(Z)$, given by
\begin{align}
    h(Z) &= \frac{c}{2\pi \mathrm{i}} \log \left[
        \frac{\mathrm{e}^{\mathrm{i}\pi/4} - \tan^{1/2}(\pi Z /2)}{\mathrm{e}^{\mathrm{i}\pi/4} + \tan^{1/2}(\pi Z /2)}
    \right].
\end{align}

Note that, by the Cauchy--Riemann relations,
\begin{align}
    \frac{\partial W_1}{\partial X} &= \frac{\partial}{\partial X}\mathrm{Im}[h(Z)] =  -\frac{\partial}{\partial Y}\mathrm{Re}[h(Z)],
\end{align}
but
\begin{align}
    -\mathrm{Re}[h(Z)] &= -\frac{c}{2\pi} \mathrm{Arg}\left[
        \frac{-\tan^{1/2}(\pi Z /2) + \mathrm{e}^{\mathrm{i}\pi/4}}{\tan^{1/2}(\pi Z /2) + \mathrm{e}^{\mathrm{i}\pi/4}}
    \right],   
\end{align}
where $\mathrm{Arg}(z)\in (-\pi,\pi]$ is the principal argument. Restricting to the fin surface, $Z = -\mathrm{i}\tilde{Y}$ with $\tilde{Y}<0$,
\begin{align}
    \left.-\mathrm{Re}[h(Z)]\right|_{X=0} &= -\frac{c}{2\pi} \mathrm{Arg}\left[
        \frac{1-\tanh(\pi \tilde{Y} /2)  +2 \mathrm{i} \tanh^{1/2}(\pi \tilde{Y} /2)}{\tanh(\pi \tilde{Y} /2) + 1}
    \right].   
\end{align}
To evaluate the argument, we note that the term in square brackets always lies in the first quadrant, and hence
\begin{align}
    \left.-\mathrm{Re}[h(Z)]\right|_{X=0} &= -\frac{c}{2\pi}
    \tan^{-1}\left( \frac{2 \tanh^{1/2}(\pi \tilde{Y} /2)}{1 - \tanh(\pi \tilde{Y} /2)}\right).   
\end{align}
Taking $\partial/ \partial Y = -\partial/ \partial \tilde{Y}$ of the above, we arrive at
\begin{align}
    \left.\frac{\partial W_1}{\partial X}\right|_{X=0} &= \frac{c/2}{\sqrt{\mathrm{e}^{2\pi \tilde{Y}}-1}} = \frac{c/2}{\sqrt{\mathrm{e}^{-2\pi Y}-1}} \quad \text{for }Y<0.
\end{align}
Finally, from (\ref{eq:Phi_1_solution}), the $O(\varepsilon)$ heat flux out of the fin follows via
\begin{align}
    -\left.\frac{\partial \Phi_1}{\partial X}\right|_{X=0} &= \frac{2(1+c)}{c} \left.\frac{\partial W_1}{\partial X}\right|_{X=0} = \frac{1 + c}{\sqrt{\mathrm{e}^{-2\pi Y}-1}} \quad \text{for }Y<0.
\end{align}

\section{Second order in the tip and fin regions}
\label{sec:second_order_tip_fin_regions}

\subsubsection*{Second order heat flux from the fin tips}
The second order temperature $\Phi_2$ in the tip region, although useful, is likely significantly more difficult to find than $\Phi_1$. This is due to a nonuniform Dirichlet condition appearing on the fin surface (see below). Consequently, we do not attempt to calculate here the local heat flux correction on the fin, $(\partial \Phi_2 / \partial X)|_{X=0}$. However, the total heat flux leaving the fin in the tip region, i.e. $\int_{-\infty}^0 (\partial \Phi_2 / \partial X)|_{X=0}\,\mathrm{d}Y$, is desirable and can be easily extracted without detailed knowledge of $\Phi_2$\footnote{For the case of infinitely conducting fins, $\Omega=\infty$, the $\Phi_2$ problem reduces to one similar to that of $W_0$, and hence can be shown to be identically zero, $\Phi_2 \equiv 0$.}. The problem for $\Phi_2$ follows from (\ref{eq:Phi_eq}), (\ref{eq:Phi_BC_1}), (\ref{eq:Phi_f_contin}) as
\begin{align}
    \nabla_{XY}^2\Phi_2 &= 0\quad \text{in } \mathcal{D}_{\mathrm{tip}}, \label{eq:Phi_2_eq} \\
    \frac{\partial \Phi_2}{\partial X} &= 0  \qquad \text{on } X = 0,1, \quad 0< Y, \\
    \Phi_2 &= \Phi_{\mathrm{f}2}  \qquad \text{on } X = 0,1, \quad Y < 0. 
\end{align}
along with (\ref{eq:Phi_f_2_solution}) determining $\Phi_{\mathrm{f}2}$. We express the matching conditions in terms of heat fluxes. The heat flux entering the gap region is completely satisfied at leading order, (\ref{eq:phi_0_heat_flux}), hence all higher orders must have no heat flux, giving
\begin{align}
    \frac{\partial \Phi_2}{\partial Y} &\to 0 \quad \text{as } Y \to \infty.
\end{align}
The heat flux in the fluid leaving the top of the fin region ($y \to 1^-$) is given by $\partial \tilde{\phi}/\partial y = \varepsilon \Delta_1 + \cdots$, resulting in the matching condition for $\Phi_2$:
\begin{align}
    \frac{\partial \Phi_2}{\partial Y} &\to \Delta_1 \quad \text{as } Y \to -\infty. \label{eq:Phi_2_matching_down}
\end{align}
As $\Phi_2$ is harmonic with known Neumann conditions along the entire boundary of $\mathcal{D}_{\mathrm{tip}}$ except the fins, we may integrate (\ref{eq:Phi_2_eq}) over $\mathcal{D}_{\mathrm{tip}}$ and apply the divergence theorem. The only nonzero contributions from the boundary are from the fins, and the fluid as $Y\to -\infty$:
\begin{align}
    2\int_0^{-\infty}\left.\frac{\partial \Phi_2}{\partial X}\right|_{X=0}\,\mathrm{d}Y + \int_0^1\left.-\frac{\partial \Phi_2}{\partial Y}\right|_{Y\to -\infty}\,\mathrm{d}X &=0.
\end{align}
The second term is known from (\ref{eq:Phi_2_matching_down}), thus the total heat flux correction from the fin in the tip region is
\begin{align}
    -\int_{-\infty}^0\left.\frac{\partial \Phi_2}{\partial X}\right|_{X=0}\,\mathrm{d}Y  &= \frac{\Delta_1}{2} = -\frac{1+c}{4\Omega}. \label{eq:total_Phi_2_heat_flux}
\end{align}

\subsubsection*{Second order temperature in the fin region: $0<y<1$}
We have found the leading order solution $\tilde{\phi}=\varepsilon \Delta_1 y + \cdots$ in the fin region ($0<y<1$), but it is now straightforward to calculate the correction to this at $O(\varepsilon^2)$. Looking at $O(\varepsilon^1)$ in (\ref{eq:phi_scaled_eq}) and $O(\varepsilon^3)$ in (\ref{eq:phi_fin_scaled_eq_1}), identical arguments to at lower orders gives that $\tilde{\phi}_3(y)$ depends only on $y$, and $\tilde{\phi}_2(y)$ is linear in $y$:
\begin{align}
    \tilde{\phi}_2(y) &= \phi_{\mathrm{f}2}(y) =  \Delta_2 y,\qquad \text{for }0<y<1,
\end{align}
where $\Delta_2$ is a constant, which we determine via an integration argument in the tip region, similar to how $\Delta_1$ was determined at the previous order. First, substituting $y=1+\varepsilon Y$ into the fin temperature $\phi_{\mathrm{f}} = \varepsilon \Delta_1 y +\varepsilon^2 \Delta_2 y + \cdots$ and taking $\mathrm{d}/\mathrm{d}Y$, we find
\begin{align}
    \frac{\mathrm{d}\phi_{\mathrm{f}}} {\mathrm{d}Y} &= \varepsilon^2 \Delta_1 + \varepsilon^3 \Delta_2 + \cdots, 
\end{align}
which, matching with the fin temperature in the tip region as $Y\to -\infty$, gives the condition at \emph{third order}
\begin{align}
    \frac{\mathrm{d}\Phi_{\mathrm{f}3}} {\mathrm{d}Y} &\to \Delta_2 \quad \text{as }Y\to -\infty. \label{eq:Phi_f_3_matching_down}
\end{align}
The equation for $\Phi_{\mathrm{f}3}$ is given by (\ref{eq:Phi_f_eq}) at $O(\varepsilon^2)$, and integrating it from $Y=0$ to $Y=-\infty$, applying the no-flux condition at the tip $Y=0$,
\begin{align}
    \lim_{Y\to -\infty} \Omega\frac{\mathrm{d}\Phi_{\mathrm{f}3}}{\mathrm{d}Y} &= \int_{-\infty}^0\left.\frac{\partial \Phi_2}{\partial X}\right|_{X=0}\,\mathrm{d}Y.
\end{align}
But the limit on the left hand side equals $\Delta_2$ from matching, (\ref{eq:Phi_f_3_matching_down}), and the right hand side was already determined to be $-\Delta_1/2$ in (\ref{eq:total_Phi_2_heat_flux}). Hence,
\begin{align}
    \Delta_2 &= -\frac{\Delta_1}{2\Omega} = \frac{1+c}{4\Omega^2},
\end{align}
and the correction in the fin region is finally
\begin{align}
    \tilde{\phi}_2(y) &= \phi_{\mathrm{f}2}(y) = \frac{1+c}{4\Omega^2}y.
\end{align}

\subsubsection*{Second order composite fin temperature}
\begin{figure}
    \centering
    \includegraphics[width=0.5\textwidth]{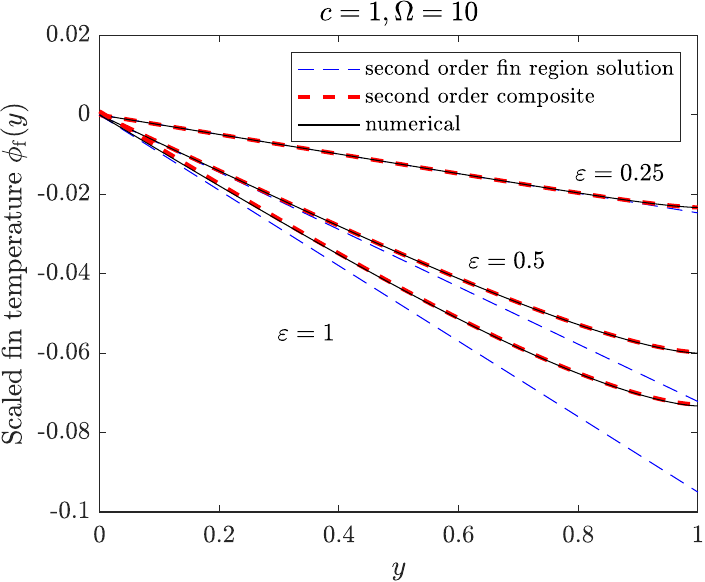}
    \caption{Comparison of the composite solution for the (scaled) temperature in the fin, given by (\ref{eq:second_order_fin_composite}), with the numerical solution.}
    \label{fig:second_order_fin_composite}
\end{figure}

We can now determine a composite solution for the temperature throughout the fin, accurate to $O(\varepsilon^2)$. Integrating (\ref{eq:Phi_f_2_solution}) from $Y=0$ (the tip) to $Y<0$,
\begin{align}
    \Phi_{\mathrm{f}2}(Y) &= \Phi_{\mathrm{f}2}(0) + \frac{(1+c)}{\pi\Omega}\int_Y^0 \tan^{-1}\left(\sqrt{\mathrm{e}^{-2\pi Y'}-1}\right)\mathrm{d}Y', \label{eq:Phi_f_2_full_solution} 
\end{align}
which can be shown to have asymptotic behaviour as $Y \to -\infty$,
\begin{align}
    \Phi_{\mathrm{f}2}(Y) &\sim \Phi_{\mathrm{f}2}(0) - \frac{(1+c)}{2\Omega}\left(Y + \frac{\log 2}{\pi}\right) +o(1).  
\end{align}
Matching with the fin region solution $\phi_\mathrm{f} = \varepsilon \Delta_1 + \varepsilon^2 (\Delta_1 Y + \Delta_2) + \cdots$ as $Y \to 0$ implies
\begin{align}
    \Phi_{\mathrm{f}2}(0) - \frac{(1+c)}{2\Omega} \frac{\log 2}{\pi} & = \frac{1+c}{4\Omega^2}, 
\end{align}
determining $\Phi_{\mathrm{f}2}(0)$. Then a second order tip--fin composite ($0\leq y\leq1$) is given by
\begin{align}
    \phi_\mathrm{f,comp} &= \varepsilon\Phi_{\mathrm{f}1} + \varepsilon^2\Phi_{\mathrm{f}2}(Y) + \varepsilon \phi_\mathrm{f1}(y) + \varepsilon^2 \phi_\mathrm{f2}(y) + \underbrace{\frac{\varepsilon(1+c)}{2\Omega}(1+\varepsilon Y) - \frac{\varepsilon^2(1+c)}{4\Omega^2}}_{-\text{overlap}}, \\
        &= -(1+c)\left[\frac{\varepsilon}{2\Omega} - \frac{\varepsilon^2 y}{4\Omega^2} - \frac{\varepsilon^2}{2\Omega}\left(\frac{\log 2}{\pi} + \frac{2}{\pi}\int_{(y-1)/\varepsilon}^0 \tan^{-1}\sqrt{\mathrm{e}^{-2\pi Y'}-1}~\mathrm{d}Y'\right) \right]. \label{eq:second_order_fin_composite}
\end{align}
A comparison of $\phi_\mathrm{f,comp}$ and the numerical solution is shown in Fig. \ref{fig:second_order_fin_composite}, and the deviation of the temperature from linear is captured well.

\section{Solution method for $\hat{\lambda}_1$ and $\hat{\phi}_1$}
\label{sec:solution_for_lambda_hat_1}
The problem for the correction in the gap region, $1 < y < 1 + c$ (or $0 < \hat{y} < 1$
in transformed coordinates) is given by (\ref{eq:phi_1_hat_eq})-(\ref{eq:lambda_phi_1_hat_eq}). After substituting $\hat{\lambda}_1$, given by (\ref{eq:lambda_phi_1_hat_eq}), into the $\hat{\phi}_1$ equation (\ref{eq:phi_1_hat_eq}), and rearranging slightly, the problem can be written in the form
\begin{align}
    \mathcal{L}_1[\hat{\phi}_1] &= \hat{\lambda}_0 \widehat{\mathcal{W}}_1 \hat{\phi}_0 - \hat{\lambda}_0^2\widehat{\mathcal{W}}_0 \hat{\phi}_0\int_0^1 \widehat{\mathcal{W}}_1 \hat{\phi}_0\,\mathrm{d}\hat{y} \quad \text{in } 0<\hat{y}<1, \label{eq:L_phi_1_hat_eq}\\
    \frac{\mathrm{d}\hat{\phi}_1}{\mathrm{d} \hat{y}} &= 0  \qquad \text{on } \hat{y}=1,\\
    \hat{\phi}_1 &= -\left(\frac{\log 2}{\pi} + \frac{1}{2\Omega}\right)  \qquad \text{on } \hat{y}=0, \label{eq:L_phi_1_hat_BC_2}
\end{align}
where the linear (integro-differential) operator $\mathcal{L}_1$ operating on $\hat{\phi}_1$ is
\begin{align}
    \mathcal{L}_1[\phi] &=
   \frac{\mathrm{d}^2 \phi}{\mathrm{d} \hat{y}^2} - \hat{\lambda}_0 \widehat{\mathcal{W}}_0 \phi + \hat{\lambda}_0^2{\widehat{\mathcal{W}}_0 \hat{\phi}_0}\int_0^1 \widehat{\mathcal{W}}_0 \phi\,\mathrm{d}\hat{y}.
\end{align}
and the forcing appearing on the right hand side of (\ref{eq:L_phi_1_hat_eq}) is known. This linear problem has two inhomogeneities: one in the governing equation (\ref{eq:L_phi_1_hat_eq}) and one in the boundary condition (\ref{eq:L_phi_1_hat_BC_2}) on $\hat{y}=0$. Hence, we can satisfy each inhomogeneity separately and linearly superimpose the corresponding solutions. In particular, if we find a $\phi^A$ that satisfies the inhomogeneous (normalised) boundary condition but homogeneous equation, given by
\begin{align}
    \mathcal{L}_1[\phi^A] &= 0, \quad \text{in } 0<\hat{y}<1,\label{eq:phi_A_eq}\\
    \frac{\mathrm{d}\phi^A}{\mathrm{d} \hat{y}} &= 0  \qquad \text{on } \hat{y}=1,\\
    \phi^A &= -1  \qquad \text{on } \hat{y}=0, \label{eq:phi_A_BC_2}
\end{align}
and a $\phi^B$ that satisfies the homogeneous boundary condition but inhomogeneous equation, given by
\begin{align}
    \mathcal{L}_1[\phi^B] &= \hat{\lambda}_0 \widehat{\mathcal{W}}_1 \hat{\phi}_0 - \hat{\lambda}_0^2\widehat{\mathcal{W}}_0 \hat{\phi}_0\int_0^1 \widehat{\mathcal{W}}_1 \hat{\phi}_0\,\mathrm{d}\hat{y}, \quad \text{in } 0<\hat{y}<1,\label{eq:phi_B_eq}\\
    \frac{\mathrm{d}\phi^B}{\mathrm{d} \hat{y}} &= 0  \qquad \text{on } \hat{y}=1,\\
    \phi^B &= 0  \qquad \text{on } \hat{y}=0, \label{eq:phi_B_BC_2}
\end{align}
then the solution for $\hat{\phi}_1$ we desire is
\begin{align}
    \hat{\phi}_1 &= \left(\frac{\log 2}{\pi} + \frac{1}{2\Omega}\right)\phi^A + \phi^B,
\end{align}
The solution for $\hat{\lambda}_1$ follows from (\ref{eq:lambda_phi_1_hat_eq}),
\begin{align}
    \hat{\lambda}_1 &= -\hat{\lambda}_0^2\left[ \left(\frac{\log 2}{\pi} + \frac{1}{2\Omega}\right) \int_0^1 \widehat{\mathcal{W}}_0 \phi^A\,\mathrm{d}\hat{y} + \int_0^1 \widehat{\mathcal{W}}_0 \phi^B\,\mathrm{d}\hat{y} + \int_0^1 \widehat{\mathcal{W}}_1 \hat{\phi}_0\,\mathrm{d}\hat{y} \right],
\end{align}
or after rearranging,
\begin{align}
    \hat{\lambda}_1 &= b_0 + \frac{b_1}{2\Omega},
\end{align}
where
\begin{align}
    b_0 &= -\hat{\lambda}_0^2\left[ \frac{\log 2}{\pi}\int_0^1 \widehat{\mathcal{W}}_0 \phi^A\,\mathrm{d}\hat{y} + \int_0^1 \widehat{\mathcal{W}}_0 \phi^B\,\mathrm{d}\hat{y} + \int_0^1 \widehat{\mathcal{W}}_1 \hat{\phi}_0\,\mathrm{d}\hat{y} \right], \\
    b_1 &= -\hat{\lambda}_0^2 \int_0^1 \widehat{\mathcal{W}}_0 \phi^A\,\mathrm{d}\hat{y}.
\end{align}
Given the solutions $\phi^A$, $\phi^B$ and the leading order solution $\hat{\lambda}_0$, $\hat{\phi}_0$, the above quantities $b_0$, $b_1$ are easily computed and have no parameters, therefore they are simply numerical constants that only need to be computed once. To do so, we use the same Chebyshev collocation methods used to compute $\hat{\lambda}_0$, but here note that the problems for $\phi^A$ and $\phi^B$ are both linear and iteration is unnecessary, with $b_0$, $b_1$ computed afterwards. Mesh refinement was performed by doubling the number of grid points until the relative change in $b_0$ and $b_1$ were both less than $10^{-4}$. The final values are given by (\ref{eq:b_0_and_b_1_values}).

\subsubsection*{Funding} TLK was supported in part by a Chapman Fellowship in the Department of Mathematics, Imperial College London. MH is supported by CBET-EPSRC Grant EP/V062298/1.

\bibliographystyle{unsrt}
\bibliography{fluids_and_heat_trans_refs_v1,References2}
\end{document}